\documentclass{aa}  

\usepackage{natbib}
\bibpunct{(}{)}{;}{a}{}{,}
\usepackage{graphicx}
\usepackage{adjustbox}
\usepackage{txfonts}
\usepackage[dvipsnames]{xcolor}
\usepackage{tabularx,multirow,booktabs,blindtext}

\usepackage[colorlinks=true,linkcolor=black, urlcolor=black, citecolor=black]{hyperref}
\hypersetup{citecolor=blue}
\hypersetup{linkcolor=blue}
\usepackage{subfigure} 
\usepackage{booktabs, caption} 
\usepackage{multicol}

\begin{document}

\title{Is planetary inward migration responsible for \object{GJ 504}'s fast rotation and bright X-ray luminosity?}
\subtitle{New constraints from eROSITA}

\author{C. Pezzotti\inst{1,2}  \and G. Buldgen\inst{1} \and E. Magaudda\inst{3} \and M. Farnir \inst{1} \and V. Van Grootel \inst{1} \and S. Bellotti\inst{4,5} \and K. Poppenhaeger\inst{6,7} }

\institute{STAR Institute, Université de Liège, Liège, Belgium:
\email{camilla.pezzotti@uliege.be}
\and Istituto Nazionale di Astrofisica - Osservatorio Astronomico di Roma, Via Frascati 33, I-00040, Monteporzio
Catone, Italy
\and Institut für Astronomie und Astrophysik, Eberhard-Karls Universität Tübingen, Sand 1, 72076 Tübingen, Germany
\and Leiden Observatory, Leiden University, PO Box 9513, 2300 RA Leiden, The Netherlands
\and Institut de Recherche en Astrophysique et Planétologie, Université de Toulouse, CNRS, IRAP/UMR 5277, 14 avenue Edouard Belin, F-31400, Toulouse, France
\and Leibniz Institute for Astrophysics Potsdam (AIP), An der Sternwarte 16, 14482 Potsdam, Germany
\and Institute for Physics and Astronomy, University of Potsdam, Karl-Liebknecht-Str. 24/25, 14476 Potsdam-Golm, Germany\\
}



\date{Received ...; accepted ...}


 \abstract
   {The discovery of an increasing variety of exoplanets in very close orbits around their host stars raised many questions about how stars and planets interact, and to which extent host stars' properties may be influenced by the presence of close-by companions. Understanding how the evolution of stars is impacted by the interactions with their planets is indeed fundamental to disentangle their intrinsic evolution from Star-Planet Interactions (SPI)-induced phenomena. In this context, GJ 504 is a promising candidate for a star that underwent strong SPI. Its unusually short rotational period ($\rm P_{rot} \sim 3.4$ days), while being in contrast with what is expected by single-star models, could result from the inward migration of a close-by, massive companion ($\rm M_{pl} \geq 2 ~M_{J}$), pushed towards its host by the action of tides. Moreover, its brighter emission in the X-ray luminosity may hint to a rejuvenation of the dynamo process sustaining the stellar magnetic field, consequent to the SPI induced spin-up.}
   {We aim to study the evolution of GJ 504 and establish whether by invoking the engulfment of a planetary companion we can better reproduce its rotational period and X-ray luminosity.}
   {We simulate the past evolution of the star by assuming two different scenarios: ``Star without close-by planet'' and ``Star with close-by planet''. In the second scenario, we investigate by means of our SPI code how the inward migration and eventual engulfment of a giant planet driven by stellar tides may spin-up the stellar surface and rejuvenate its dynamo. We compare our theoretical tracks with archival rotational period and X-ray data of GJ 504, collected from the all-sky surveys of the ROentgen Survey with an Imaging Telescope Array (eROSITA) on board the Russian Spektrum-Roentgen-Gamma mission (SRG).}
   {Despite the large uncertainty on the stellar age, we found that the second evolutionary scenario characterised by the inward migration of a massive planetary companion is in better agreement with the short rotational period and the bright X-ray luminosity of GJ 504, thus strongly favouring the inward migration scenario over the one in which close-by planets have no tidal impact on the star.
}
   {}

\keywords{Planet-star interaction - Planetary systems - Stars: evolution - Stars: rotation - Stars: solar-type}

\titlerunning{}
\maketitle

\section{Introduction}

With the discovery of thousands of exoplanets in the last few decades (5678 as of June 2024, Nasa Exoplanet Archive\footnote{https://exoplanetarchive.ipac.caltech.edu/}), the need to precisely characterise their host stars has become crucial in order to accurately derive the properties of the systems \citep[e.g.,][]{Adibekyan2018}, together with understanding their formation and evolution. Since their birth, stars shape the evolution of exoplanetary systems. At the same time, host stars’s properties may be significantly influenced by the presence of interacting close-by planetary companions, potentially resulting in anomalous rotational periods \citep[e.g.,][]{Privitera2016B, Ilic2024} and magnetic activity, with brighter X-ray luminosity emissions \citep[e.g.,][]{Shkolnik2003,Poppenhaeger2014, Pillitteri2022, Ilic2023} (among others). Understanding how the evolution of stars is impacted by the interactions with their planets is thus fundamental to disentangle their intrinsic evolution from SPI-induced phenomena on the one hand, and to help determining properties that could not be derived otherwise (for example trace back their rotational history), on the other one. 

In this context, GJ 504 (a.k.a. HD 115383,  TIC 397587084) is a promising candidate for a star that underwent strong star-planet interactions. GJ 504 is an isolated G0 spectral type star, slightly more massive than the Sun ($\rm M_{\star} \sim 1.22~M_{\odot}$), that hosts a directly-imaged substellar companion at a projected distance of $\rm \sim 43.5~AU$ \citep{Kuzuhara2013}. Because of the significant uncertainties in determining GJ 504's evolutionary state, with age estimations varying from hundreds of Myr to several Gyr \citep{Valenti2005,  Takeda2007, Holmberg2009, daSilva2012, Kuzuhara2013, Fuhrmann2015, DOrazi2017, DiMauro2022}, the nature of the companion is still hotly debated, and very different values for its mass have been proposed in literature ($\rm  1 \lesssim M_{c}/M_{J} \lesssim 25 $) \citep{Kuzuhara2013, Fuhrmann2015}. 

Establishing the age of isolated stellar objects is particularly hard. Different methods could be used \citep{Soderblom2014}, including gyrochronology and activity indicators, for the age-rotation-activity relation \citep{Skumanich1972, Barnes2007}, or comparison of classical spectroscopic parameters with model isochrones \citep[e.g.][]{Pont2004}. \citet{Kuzuhara2013} estimated the age of GJ 504 by means of the several methods, but considered the one derived from gyrochronology and activity indicators, based on $\rm P_{rot} = 3.329~days$ \citep{Donahue1996,Messina2003} and chromospheric activity indices (Ca II H and K lines, $\rm log (R_{HK}^{\prime}) = -4.45$), as the most likely one, with value $\rm t = 160^{+350}_{-60}~Myr$. A few years later, \citet{Fuhrmann2015} carried out a detailed analysis on high-resolution, high-quality spectra, that led to a major revision of the stellar gravity parameter $\rm log(g) = 4.23 \pm 0.10$, about $\rm 8\%$ less than the value in \citet{Kuzuhara2013}. From the isochrones fitting method, they found an age much closer to the one of the Sun ($\rm t = 4.5^{+2}_{-1}~Gyr$) \citep{Fuhrmann2015}. To explain the short rotational period, intense chromospheric activity and bright X-ray luminosity ($\rm Log(L_{X}/L_{Bol}) = -4.42$, \cite{Voges1999,Wright2011}), and reconcile these indicators with the isochronal age, \citet{Fuhrmann2015} invoked the engulfment\footnote{With the term ``engulfment'' in this work we refer to the destruction of the substellar companion at the Roche limit.} of a close-by planetary companion, that would have spun-up the stellar surface and enhanced its activity levels \citep{Oetjens2020}. \citet{DOrazi2017} reassessed the fundamental properties of GJ 504, finding that a comparison of their spectroscopic parameters with isochrones provided an age between 1.8 and 3.5 Gyr, with most probable value $t \rm \approx 2.5~Gyr$. They also envisaged a possible engulfment scenario to reconcile the different age indicators, and tested this hypothesis by means of a tidal evolution code. They found that the engulfment of a hot Jupiter, with initial mass not larger than $\rm \approx 3~M_{J}$ and initial orbital distance $\rm \approx 0.03~AU$, could be a very likely scenario. \citet{Bonnefoy2018} revisited the system by means of high-contrast imaging, interferometric and radial velocity observations. From their analysis, they retrieved an interferometric radius for the host star of $\rm R = (1.35 \pm 0.04)~R_{\odot}$, that is compatible with two isochronal ages $\rm (21 \pm 2)~Myr$ and $\rm (4.0 \pm 1.8)~Gyr$. The mass of the substellar companion would correspond to $\rm 1.3^{+0.6}_{-0.3}~M_{J}$ or $\rm 23^{+10}_{9}~M_{J}$, respectively. The authors also revised the almost pole-on line-of-sight stellar rotation axis inclination, which is $\rm 18.6^{+4.3}_{-3.8}$ degrees or $162.4^{+3.8}_{-4.3}$ degrees. They also excluded the presence of additional objects (with $\rm 90\%$ probability) more massive than $\rm 2.5$ and $\rm 30~M_{J}$ with semi-major axes in the range $\rm 0.01-80~AU$ for the young and old isochronal ages, respectively. More recently, \citet{DiMauro2022} attempted to employ asteroseismic techniques on the observational data collected by the Transiting Exoplanet Survey Satellite space mission \citep[TESS,][]{Ricker2014} to accurately characterise GJ 504. Unfortunately, the non-detection of solar-like oscillations, that \citet{DiMauro2022} ascribed to the high level of the stellar magnetic activity, hindered their analysis. Nevertheless, the results deduced by TESS photometric data, supported by the Mount Wilson Observatory long term campaign spanning nearly 30 years, and by modelling procedures, allowed the authors to refine GJ 504 fundamental parameters. Among these, they derived an age $\rm \leq 2.6~Gyr$, which is in agreement within the quoted uncertainties with previous findings by \citet{Kuzuhara2013} and \citet{DOrazi2017}. From the TESS light curves, they also identified a clear modulation corresponding to a rotational period $\rm P_{rot}  = 3.4~days$, which confirms the average value used in \citet{Kuzuhara2013}. Finally, from Mount Wilson Observatory data they obtained the detection of a main magnetic cycle of 11.97 years, that together with the relatively short rotational period, locate GJ 504 before the transition to the weakened magnetic braking regime, as theorised by \citet{VanSaders2016}, during which the large scale magnetic field would fail to efficiently brake the stellar surface.

The aim of our work is to study the peculiar properties of GJ 504 by testing the impact of a putative, close-by planet's inward migration on the evolution of stellar rotational period and X-ray luminosity. To this purpose, we first look for the optimal stellar parameters representative of GJ 504, by means of a minimisation procedure based on classical spectroscopic/interferometric parameters, and correspondingly compute best-fit stellar models (see Sect.~\ref{Sec:Opt}). Subsequently, we couple the stellar models to our SPI code (see Sec.~\ref{Sec:SPIcode}), in which the evolution of the stellar surface rotation rate and X-ray luminosity is computed, simultaneously to the evolution of the orbit of a close-by planetary companion, driven by the dissipation of tides within the host star (see Sect.~\ref{Sec:SPI}). We envisage two evolutionary scenarios:
\begin{itemize}

\item ``Star without close-by planet'' (Sect.~\ref{Sub: Single Star}), in which no massive, no close-by companions affect the evolution of the host star;

\item ``Star with close-by planet'' (Sect.~\ref{Sub: close-by planet}), in which a putative, close-by planet strongly impacts the evolution of its host star \footnote{The detected companion at $\rm \approx  43~AU$ is thus considered to have negligible impact on the evolution of the host star}. 

\end{itemize}  

We thus compare the results of the simulations with observational data for the rotational period and X-ray luminosity. Noticeably for the latter one, in addition to the ROSAT \citep{Trumper1982} data, we perform a detailed analysis of the X-ray data taken with the ROentgen Survey with an Imaging Telescope Array \citep[eROSITA;][]{Predehl2021} on board the Russian Spektrum-Roentgen-Gamma mission \citep[SRG;][]{Sunyaev2021} (see Sect.~\ref{Sec:eROSITA}), and compare with previous results found in \citet{Foster2022}. We searched for an X-ray counterpart of GJ\,504 among the sources listed in the five all-sky surveys of eROSITA \citep[eRASS;][]{Merloni24.0}. The first four surveys are completed and count six months of observation, while the last survey was suspended after almost two months because the science operations of the instrument were paused. Finally, in Sect.~\ref{Sec:Conclusion} we draw our conclusions.

\section{Method and physics}
\label{Sec:MethodTools}

To carry out our study on GJ 504, we proceed according to the following steps: firstly, we derive the optimal parameters for the star (initial mass, radius, chemical composition, ...) by means of a minimisation technique based on classical spectroscopic/interferometric parameters (unfortunately, asteroseismic indicators are not available for this star); secondly, best-fit stellar models are computed based upon the optimal parameters found before; finally, the stellar models are coupled to our SPI code in which the evolution of the star is computed by envisaging two potential scenarios: ``Star without close-by planet'', or ``Star with close-by planet''.

\subsection{Optimal parameters' search and stellar model}
\label{Sec:Opt}

To compute stellar models for GJ 504, we started by deriving optimal stellar parameters by means of a two-steps, global and local minimisation procedure. For the global minimisation, we employed the SPInS software \citep{Lebreton2020}, which is based on a Monte Carlo Markov Chain (MCMC) approach and Bayesian statistics to derive probability distribution functions for stellar parameters. For the local minimisation instead, we used the Levenberg-Marquardt algorithm \citep[e.g.][]{Miglio2005, Farnir2020}.

In both minimisation procedures, we used models computed with the Liège stellar evolution code (CLES) \citep[e.g.][]{Scuflaire2008}. For the set-up of input physics, we considered FreeEOS \citep{Irwin2012} equation of state, AGSS09 \citep{Asplund2009} abundances, and OPAL \citep{IglesiasRogers1996} opacity tables for solar mixture. The classical mixing-length theory was applied for convection, with a solar calibrated value $\rm \alpha_{MLT} = 2.01$. For the outer boundary conditions, we used \citet{Vernazza1981}.    

In this context, we aimed at determining the values of four free parameters ($\rm M_{\star}$, age, $\rm X_{0}$ and $\rm Z_{0}$), by using four observational constraints taken from \citet{DiMauro2022} (see references therein): $\rm Teff = (6205 \pm 20)~K$, $\rm R = (1.35 \pm 0.04)~R_{\odot}$, $\rm [Fe/H] = 0.22 \pm 0.04$, and $\rm log(g) = 4.29 \pm 0.07$. 

The results obtained from the first step modelling with SPInS are: $\rm M_{\star} = (1.30 \pm 0.05)~M_{\odot}$,  $\rm X_{0} = 0.70 \pm 0.02$, $\rm Z_{0} = 0.025 \pm 0.002$, and age$\rm = (1.81 \pm 0.46) $ Gyr. While this solution is limited to the fixed input physics used for the computation of the models grid, in the second step  the local minimisation procedure allows to further explore the parameter space thanks to the use of the Levenberg-Marquardt algorithm. With the computation of models on the fly, it is possible to investigate the impact of changing the input physics on the stellar parameters. In particular, we tested the effect of including a moderate amount of overshooting ($ \alpha_{\rm Ov} = 0.1~H_{\rm p}$, with $H_{\rm p}$ being the pressure scale height). We also changed the outer boundary conditions, using the Eddington $T(\tau)$ relations, and the corresponding solar calibrated value for the mixing length $\rm \alpha_{MLT} = 1.8$. The optimal parameters derived by means of this procedure are: $\rm M_{\star} = (1.29 \pm 0.21)~M_{\odot}$,  $\rm X_{0} = 0.70 \pm 0.09$, $\rm Z_{0} = 0.025 \pm 0.002$, and  age$\rm = (2.11 \pm 1.75) $ Gyr. 

Both solutions agree very well on the mean values, but from the local minimisation procedure we derived larger uncertainties, especially for the age. This is due not only to the broader input physics used in the Levenberg-Marquardt algorithm, but also to the intrinsic differences in comparison to SPInS for the computation of the uncertainties. 

In general, these solutions appear to be in good agreement with the ones found in \citet{DiMauro2022}, although a smaller uncertainty on the age estimation was found in our case. According to our analysis, an age as young as $\rm \approx 200~Myr$ seems to be disfavoured for this star. Nevertheless, it is worth stressing that only seismic constraints would likely allow us to derive more precise age estimates \citep{Soderblom2010}. 

\begin{figure}
\includegraphics[width=\linewidth]{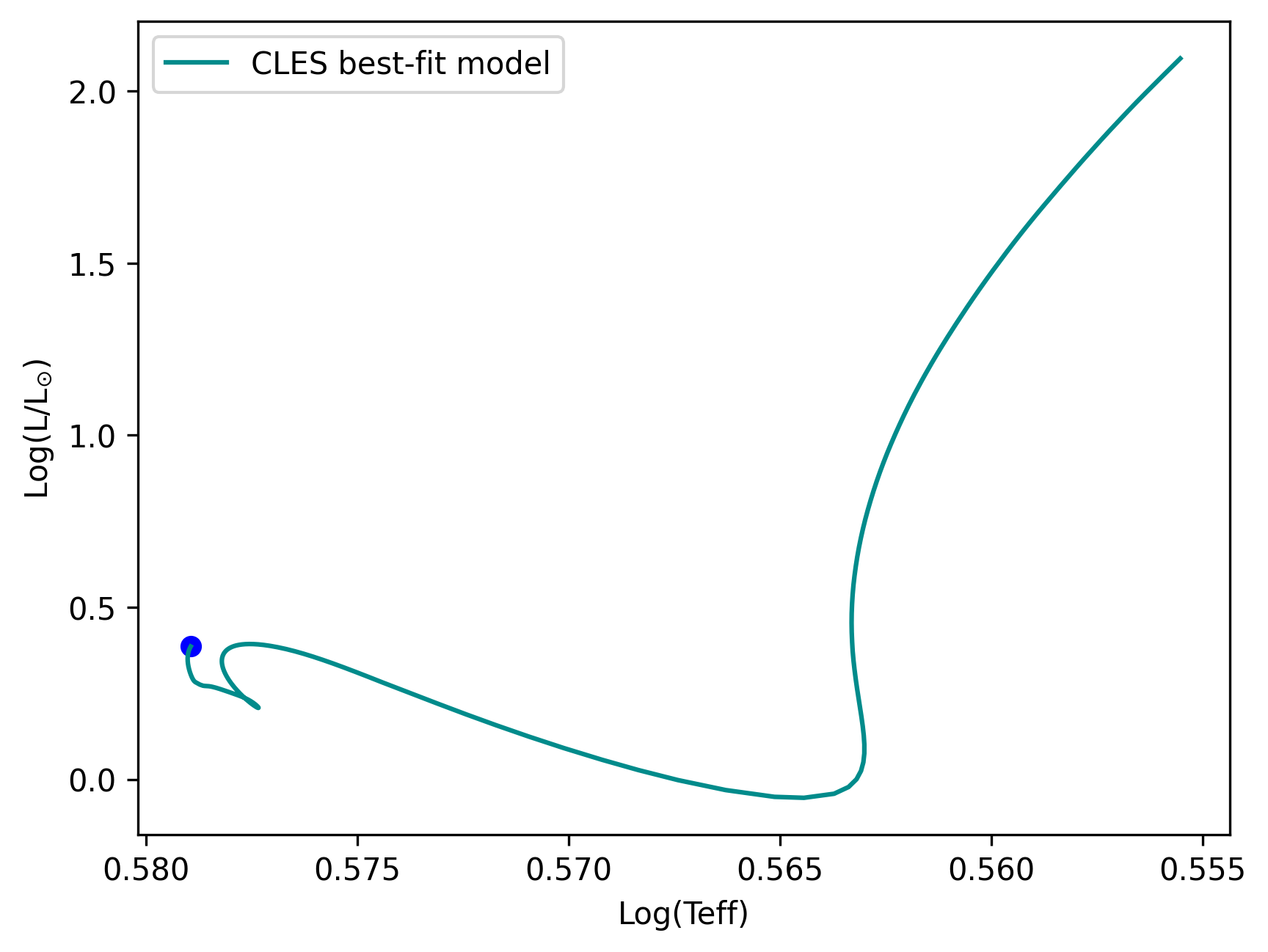}
\caption{Hertzsprung-Russell of GJ 504 as obtained from the best-fit stellar model ($\rm Teff = 6202~K$, $\rm log(g) = 4.29$, $\rm R = 1.35~R_{\odot}$, $\rm L/L_{\odot} = 2.43$) computed with CLES.}
\label{Fig: HR}
\end{figure}

We thus proceeded by computing evolutionary sequence of GJ 504 using CLES, to which the derived optimal parameters ($\rm M_{\star} = 1.29~M_{\odot}$,  $\rm X_{0} = 0.70$, $\rm Z_{0} = 0.025$) were provided as input and constraints. According to this model, GJ 504 is a Main Sequence (MS) star, with a central abundance of hydrogen $\rm X_{c} = 0.37$, mass of the convective envelope $\rm M_{env} = 2.5 \times 10^{-3}~M_{\star}$, radius of the convective envelope $\rm R_{env} = 0.18~R_{\star}$, and an inner convective core with $\rm M_{cc} = 1.3 \times 10^{-2}~M_{\star}$ and $\rm R_{cc} =4.3 \times 10^{-2}~R_{\star}$. In Fig.~\ref{Fig: HR}, we show the evolutionary track corresponding to the best-fit model. For a detailed description of the physics included in CLES, we refer the interested reader to \citet{Scuflaire2008}.

As mentioned above, the computed evolutionary sequence of GJ 504 is provided to our SPI code to study its past evolution. Given the significant uncertainties derived on GJ 504's parameters, in the following we will discuss our results by considering the broadest age interval ($\rm (0.36 - 3.86)$ Gyr), while showing for comparison both average values and their corresponding uncertainties in the figures. While at different age values would correspond different stellar masses and chemical compositions, considering the physics included in our SPI code, we can reasonably assume that our conclusion remain valid within the 4-D space ($\rm M_{\star}$, age, $\rm X_{0}$ and $\rm Z_{0}$) of variation for our best-fit stellar model.

\subsection{Star-planet interactions code}
\label{Sec:SPIcode}

The stellar models computed for GJ 504 are provided as input to our SPI code. In this code it is possible to compute the rotational evolution of the star, while accounting for two main type of star-planet interactions, which are the gravitational-tidal and radiative one \citep[e.g.][]{Cuntz2000, Vidotto2020, Strugarek2024}. While a more detailed description of the physics implemented in the code is provided in \citet{Privitera2016B,Rao2018,Pezzotti2021,Fellay2023}, in the following we just recall the equations of interest for the evolution of a putative planet orbiting closely around GJ 504.

\subsubsection{Host-star surface rotation's evolution}

The evolution of the stellar surface rotation rate is computed assuming that the host-star rotates as a solid body on the PMS and MS phases \citep{Rao2021}. Given the relative shallowness of the convective envelope characterising GJ 504, and the fact that the differential rotation in this region is less significant compared to the almost flat profile of internal rotation in the core, this is a reasonable assumption, which has been supported by seismic analyses conducted on MS, solar-like stars \citep{Garcia2007,Nielsen2015, Benomar2015, Benomar2018, Betrisey2023}, and $\rm \gamma$Dor stars \citep{Saio2021}. The equation describing the variation of the stellar angular momentum reads as:

\begin{equation}
 \dot{L}_{\star} = \dot{L}_{wb} + \dot{L}_{tides},
\end{equation}
\label{Eq:L_evol}

where $\rm \dot{L}_{wb}$ accounts for the rate of change of angular momentum due to magnetised winds \citep{Matt2015, Matt2019}, with equation:

\begin{equation}
\dot{L}_{wb} =
\begin{cases}
\rm -T_{\odot} \left(\dfrac{R_{\star}}{R_{\odot}} \right)^{3.1} \left( \dfrac{M_{\star}}{M_{\odot}} \right)^{0.5} \left(\dfrac{\tau_{conv}}{\tau_{conv \odot}} \right)^{p} \left(  \rm \dfrac{\Omega_{\star}}{\Omega_{\odot}} \right)^{p+1} &, \rm Ro >\dfrac{Ro_{\odot}}{\chi},\\
\rm -T_{\odot} \left(\dfrac{R_{\star}}{R_{\odot}} \right)^{3.1} \left( \dfrac{M_{\star}}{M_{\odot}} \right)^{0.5} \chi^{p} \left( \dfrac{\Omega_{\star}}{\Omega_{\odot}} \right) & , \rm  Ro \leq\dfrac{Ro_{\odot}}{\chi}, 
\end{cases}
\end{equation}
\label{Eq:Lwb}

with $\rm R_{\odot}$ and $\rm M_{\odot}$ the radius and mass of the Sun, $\rm R_{\star}$ and $\rm M_{\star}$ the radius and mass of the considered star, $\rm \tau_{conv}$ the convective turnover timescale (see Eq.~\ref{Eq:tauconv}), $\rm Ro$ the stellar Rossby number, defined as the ratio between the star rotational period ($\rm P_{rot}$) and $\rm \tau_{conv}$, and $\rm Ro_{\odot}$ the solar Rossby number. The braking constant $\rm T_{\odot}  = 8 \times 10^{30} ~erg$ is calibrated to reproduce the surface rotation rate of the Sun, and the coefficient $\rm p$ is taken equal to $\rm 2.1$. The quantity $\rm \chi \equiv Ro_{\odot}/Ro_{sat}$ indicates the critical rotation rate for stars with given $\rm \tau_{conv}/\tau_{conv_{\odot}}$, defining the transition from the saturated to unsaturated regime. In this work, $\rm \chi$ is considered equal to 13 \citep{Pezzotti2021}. For what concerns the braking of the stellar surface by magnetised stellar winds, it is possible to account for the onset of a ``weakened magnetic braking'' regime \citep{VanSaders2016} when the stellar Rossby number ($\rm Ro$) is equal to a critical value $\rm Ro_{cri} = 0.92~Ro_{\odot}$ \citep{Metcalfe2024}: once a star enters this regime, the efficiency of the braking at its surface gets hampered by a shift in the morphology of the magnetic field, from larger to smaller spatial scales \citep{Reville2015,Garraffo2016}, and/or by an abrupt change in the mass-loss rate \citep{OFionnagain2018}. In our code, we account for this effect by simply turning off the magnetic braking component in the computation of the surface angular momentum evolution for Rossby numbers larger than $\rm Ro_{cri}$. The component $\dot{L}_{tides}$ in Eq.\ref{Eq:L_evol} accounts for the exchange of angular momentum between the star and the planetary orbit due to tidal dissipation \citep{Rao2021}, and its equations reads as:

\begin{equation}
\dot{L}_{tides} = \rm - \left[ \frac{1}{2} M_{pl} \left(  \dfrac{\dot{a}}{a} \right)_{tides} \right] \times \sqrt{\rm G\left( M_{\star} + M_{pl}\right) a},
\end{equation}

where the term $\rm  \left( \dfrac{\dot{a}}{a}\right)_{tides}$ indicates the net variation of the planetary orbital distance ($\rm a$) due to tidal dissipation (see Eq.~\ref{Eq:eq}, \ref{Eq:DT}), $\rm M_{pl}$ is the planetary mass, and $\rm G$ is the universal gravitational constant.

\subsubsection{Tidal interaction}

In our code, star-planet interactions related with the dissipation of equilibrium/dynamical tides within the stellar convective envelope are considered, which can lead to planetary migration through the transfer of angular momentum from the planetary orbit to the stellar surface. In the hypothesis that the planet is on a circular, coplanar orbit around the host star, the net change of its orbital distance can be written as:

\begin{equation}
\rm \left(\dfrac{\dot{a}}{a}\right) = - \dfrac{\dot{m}_{pl}}{m_{pl} + M_{\star}} + \left(  \dfrac{\dot{a}}{a}  \right)_{eq} + \left(  \dfrac{\dot{a}}{a}  \right)_{dyn} ,
\label{Eq:netA}
\end{equation}

where the first term is related with planetary mass loss due to photoevaporation, in the assumption that the mass is lost into space and does not contribute to the exchange of angular momentum between the star and the planetary orbit. The last two terms refer to the contribution of equilibrium and dynamical tides, respectively. Equilibrium tides are most efficiently dissipated in low-mass stars convective envelopes, due to turbulent friction induced by convection, while their damping is less efficient in other regions \citep{Zahn1966, Zahn1977}. In our approach we therefore account only for the dissipation of equilibrium tides in convective envelopes, and the equation accounting for their contribution reads as:

\begin{equation}
\rm \left(   \dfrac{\dot{a}}{a}  \right)_{eq} = \dfrac{f}{\tau_{conv}} \dfrac{M_{env}}{M_{\star}} q(1+q) \left(  \dfrac{R_{\star}}{a} \right)^{8} \left(  \dfrac{\Omega_{\star}}{\omega_{pl}} -1 \right),
\label{Eq:eq}
\end{equation}

where $\rm a$ is the orbital distance, $\rm \dot{a}$ is its derivative with respect to time, $\rm M_{env}$ is the mass contained in the stellar convective envelope, $\rm q$ is the ratio between planetary and stellar mass, $\rm \Omega_{\star}$ is the host star surface rotation rate, $\rm \omega_{pl}$ is the orbital frequency and $\rm R_{\star}$ is the stellar radius. The term $\rm \tau_{conv}$ is the convective turnover timescale, computed as in \citet{Rasio1996}:

\begin{equation}
\rm \tau_{conv} = \left(   \dfrac{M_{env} \left( R_{\star} - R_{b} \right)^{2}}{3L_{\star}}  \right)^{1/3}.
\label{Eq:tauconv}
\end{equation}

where $\rm R_{b}$ is the radius at the bottom of the convective envelope, and $\rm L_{\star}$ the bolometric luminosity of the host star. The term $\rm f$ in Eq.~\ref{Eq:eq} is a numerical factor obtained from integrating the viscous dissipation of the tidal energy across the convective zone \citep{Villaver2009}, which is $\rm f = \left(  P_{orb}/2\tau_{conv}  \right)^{2}$ \citep{Goldreich1977} when $\rm \tau_{conv} < P_{orb}/2$, otherwise $\rm f = 1$.

Contrarily to equilibrium tides, dynamical tides might be efficiently dissipated in both convective and radiative regions, depending on the stellar and planetary companion properties. In convective regions, they are excited in the form of inertial waves driven by the Coriolis force, whenever the tidal frequency ($\rm \omega_{t}$) ranges between $\rm \left[ -2\Omega_{\star}, 2\Omega_{\star}   \right]$ \citep{Ogilvie2007}. We account for the impact of dynamical tides in convective envelopes in the form of a frequency-averaged tidal dissipation of inertial waves \citep{Ogilvie2013, Mathis2015, BolmontMathis2016}. In particular, we assume a two-layer model (core-envelope), in which each of them is characterised by a uniform density \citep{Mathis2015}. If the planet is on a circular-coplanar orbit, then dynamical tides are active whenever $\rm \omega_{pl} < 2 \Omega_{\star}$ \citep{Ogilvie2007}, where $\rm \omega_{pl}$ is the orbital frequency of the planet. The equation describing the contribution of this type of tides on the evolution of the planetary orbital distance reads as:

\begin{equation}
\rm \left( \dot{a}/a \right)_{dyn} = \left( \dfrac{9}{2Q^{\prime}_{d}} \right)q \omega_{pl} \left( \frac{R_{\star}}{a} \right)^5 \dfrac{(\Omega_{\star} - \omega_{pl})}{\mid \Omega_{\star} - \omega_{pl}\mid},
\label{Eq:DT}
\end{equation}
with $\rm Q^{\prime}_{d} = 3/(2D_{\omega})$ (modified tidal dissipation factor) and  $\rm D_{\omega} = D_{0\omega}D_{1\omega}D_{2\omega}^{-2}$. The `D' terms are defined as follows

\begin{equation}
\begin{cases}
\rm D_{0\omega} = \dfrac{100\pi}{63} \epsilon^{2} \dfrac{\alpha^5}{1 - \alpha^5} (1 - \gamma)^2,\\
\rm D_{1\omega} = (1 - \alpha)^4 \left( 1 + 2\alpha + 3\alpha^2 + \frac{3}{2} \alpha^3 \right)^2 ,\\
\rm D_{2\omega} = 1 + \frac{3}{2}\gamma + \frac{5}{2 \gamma}\left( 1 + \frac{\gamma}{2} - \frac{3 \gamma^2}{2} \right) \alpha^3 - \frac{9}{4}\left(1 - \gamma\right)\alpha^5  ,
\end{cases}
\label{Eq:Diss_factors}
\end{equation}
where $\rm \alpha = R_{b}/R_{\star}$, $\rm \beta = M_{b}/M_{\star}$, $\rm \gamma = \dfrac{\alpha^3 (1 - \beta)}{\beta (1 - \alpha^3)}$, and $\rm \epsilon = \Omega_{\star}/\sqrt{GM_{\star}/{R_{\star}^3}}$. The terms $\rm M_{b}$ and $\rm R_{b}$ are the mass and the radius of the radiative core, considered as the region of the star below the base of the convective envelope. The term $\rm D_{\omega}$ is the frequency-averaged tidal dissipation \citep{Ogilvie2013}. We notice that assuming this schematic stratification for the stellar structure, together with the use of a frequency-average dissipation rate for the dynamical tide, adds a certain degree of uncertainty on our results. Nevertheless, while this approach suffers from some schematic simplifications, it has the advantage of providing us with relevant orders of magnitude of tidal dissipation rates accounting for the evolution of the structural and rotational parameters of the host star \citep{Mathis2015, BolmontMathis2016, Rao2018, Barker2020}.

In general, for the formalism treated above, the impact of tides is to widen (respectively shrink) the planetary orbit when it is beyond (respectively inside) the corotation radius, defined as the distance at which the orbital and host star rotational periods are equal, namely:
\begin{equation}
\rm a_{cor} = \left[  G \left(  M_{\star} + M_{pl} \right)/ \Omega_{\star}^2 \right]^{\frac{1}{3}}
\label{Eq:coradius}
\end{equation}

where $\rm G$ is the universal gravitational constant, $\rm M_{\star}$ is the mass of the host star. For what concerns the formalism used in this work for dynamical tides in the stellar convective envelope, these are efficiently excited in the form of inertial waves when the planetary orbital distance is larger than a minimum critical value, defined as $\rm a_{min} = 4^{-\frac{1}{3}} \times a_{cor}$ \citep{Ogilvie2007}.

In the case of early G- and late F-type stars, a convective core is present in addition to the envelope during the MS, separated by a radiative region. Nevertheless, dynamical tides excited in convective cores are weakly dissipated. In this case indeed, inertial waves would propagate in a fully convective sphere (the inner core), without the possibility to reflect and lead to the formation of shared waves attractors, that may induce strong dissipation \citep{Ogilvie2007, Ogilvie2004}. The dissipation of inertial waves in convective cores is thus not considered in our SPI code.

In general, dynamical tides might be excited also in stellar radiative regions in the form of gravity waves, and cause the migration of planets by means of thermal, (weakly) nonlinear, or resonance-locking dissipative effects \citep{Goodman1998, Barker2010, Weinberg2012, Ivanov2013, Essick2016, Fuller2017, MaFuller2021}. For solar-type stars with radiative cores, the most likely dominant mechanisms for the dissipation of tidally excited gravity waves are: nonlinear dissipation, triggered by planetary companions with mass $\rm M_{p} \gtrsim 0.3 ~M_{J}$ \citep{Essick2016}; or wave breaking, whose triggering requires the planetary mass to be larger than a critical value, which strongly varies as a function of the stellar age from $\rm \sim 10^{2}-10^{3}~M_{J}$ at $\rm t_{\star} \sim 10^{8}~yr$ to $\rm \sim 10^{-1}-10^{-2} ~M_{J}$ at $\rm t_{\star} \sim 10^{10}~yr$ for a $\rm 1~M_{\odot}$ star \citep{Barker2020,Lazovik2021}.

Stars slightly more massive than the Sun  ($\rm 1.1 \lesssim M_{\star}/M_{\odot} \lesssim 1.6$), are generally characterised by a rather complex structure on the MS, composed by an inner convective core surrounded by a radiative region, which is topped by a shallow convective envelope. In this configuration, the tidally excited gravity waves which travel from the radiative-convective interface towards the stellar center are reflected at the convective core interface, typically resulting into inefficient dissipation \citep{Barker2020}. An alternative scenario has been proposed for the efficient dissipation of gravity waves propagating in such mixed cores, which consist in the conversion of gravity waves into magnetic waves \citep{Lecoanet2017, Lecoanet2022, Rui2023}: in this context, tidally excited gravity waves travelling from the radiative-convective envelope boundary towards the stellar center, upon encountering a sufficiently strong magnetic field generated by the dynamo at work in the convective core, would be fully converted into outwardly propagating magnetic waves, and finally get dissipated by radiative or ohmic diffusion \citep[see][and references therein]{Duguid2024}. By using stellar models with initial masses between $\rm 1.2$ and $\rm 1.6~M_{\odot}$, \citet{Duguid2024} showed that the wave conversion mechanism may operate for a significant fraction (order of Gyrs) of stars MS lifetimes, whenever the local radial magnetic field in proximity of the convective core is larger than a critical value at which the radial wavenumber of tidally excited gravity waves and magnetic waves match \citep{Fuller2015, Lecoanet2017, Lecoanet2022, Rui2023}. Differently from the wave breaking mechanism, wave conversion does not require any threshold on the minimum mass of the planet to be triggered. Another promising dissipation mechanism of tidally gravity waves in this type of stars is tidal resonance-locking, in which a planet locks into resonance with a tidally excited stellar gravity mode. This process, similarly to the wave conversion one, can operate for planets of any mass in stars with convective cores \citep{MaFuller2021}.

It is worth stressing that all the tidal excitation and dissipation processes mentioned above strongly depend on the structural and rotational properties of the host star, and on its age. The significant uncertainties derived in Sect.~\ref{Sec:Opt} on the mass, age and initial chemical composition of GJ 504, essentially due to a lack of asteroseismic constraints, make it difficult to understand which tidal excitation mechanism might have dominated its past evolution, in the hypothetical presence of a close-by planetary companion. While the last two mechanisms mentioned above could have played a major role in this sense, given that GJ 504 could harbour an inner convective core according to our best-fit model, in this work we solely focus on the impact of equilibrium and dynamical tides dissipated within the stellar convective envelope and leave the investigation of the impact of tides dissipated in the radiative regions to a future work.

\subsubsection{Radiative interaction}

In the SPI code it is possible to compute the erosion of the planetary atmosphere due to the impact of stellar X-ray and extreme ultraviolet (EUV) flux \citep[e.g.,][]{Watson1981,Lammer2003,Jin2014}. The evolution of the X-ray luminosity is computed by following a recalibration of the $\rm Rx-Ro$ relationship of \citet{Johnstone2021} as in \citet{Pezzotti2021}, where $\rm Rx = L_{X}/L_{Bol}$ is the ratio between the X-ray luminosity and the bolometric one, while $\rm Ro$ is the stellar Rossby number. The $\rm L_{X}$ and $\rm P_{rot}$ evolution are tightly linked by means of the dynamo-activity-rotation feedback loop \citep[e.g.][]{Parker1955,Wilson1966,Kraft1967}, therefore it is important to simultaneously compute these two quantities. For the computation of the EUV flux, we refer to \citet{Johnstone2021}.

Depending on the planetary and system properties, the mass loss rate from the planet may be computed either by using an energy limited-like formula \citep[][]{Watson1981, Erkaev2007}, in which the factor accounting for the evaporation efficiency is estimated by following \citet{Salz2016} or \citet{Caldiroli2022}; or by interpolating/extrapolating the grid of upper atmosphere models of \citet{Kubyshkina2018, Kubyshkina2021}. For what concerns the computation of the planetary radius, depending on the initial mass, fraction of atmosphere and composition, we use fitting formulae from \citet{Lopez2014, Chen2016} or mass-radius relations \citep{Otegi2020, Bashi2017}. For the range of planetary masses considered in this work, we used \citet{Bashi2017}.

\section{Analysis of the X-ray data from eROSITA}
\label{Sec:eROSITA}

\subsection{Method and tools}
\label{subsec:MethodTools}
We looked for X-ray detections of GJ\,504 in the five all-sky surveys of eROSITA \citep[eRASS;][]{Merloni24.0}, by mean of the tool {\sc erose} provided by the German consortium. First, we retrieved the sky map where the source is located during the $5$ surveys. Then, we performed the source detection using the eSASSusers\_240410 software release \citep{Brunner22.0} and within the energy band of $0.2-5.0$\,keV, usually adopted for M~dwarfs \citep{Magaudda22.0}. We compiled the list of the detected sources for each of the $5$ surveys using the dedicated eSASS pipeline, {\sc ermldet}, for which we used a threshold detection maximum likelihood of $6.0$. 
Finally, we propagated the \textit{Gaia}-DR3 coordinates of GJ\,504 for their proper motion and at the epoch of the five eROSITA surveys and cross-matched within $10^{\prime\prime}$ these coordinates with those in the result of our source detection. The results are shown in Table~\ref{tab:xobs_log}, where we present the eROSITA survey to which the observation refers (col.1), actual observational period of GJ\,504 (col.2), separation between the proper motion corrected coordinates and the X-ray source position from eROSITA (col.3), number of source counts with the detection likelihood (cols.4\&5) and source count rate (col.6).
\begin{table*}
\begin{center}
    \captionsetup{justification=centering}
    \caption{eROSITA observation log of GJ\,504.}
    \label{tab:xobs_log}
    \begin{tabular}{cccccc}
      \midrule[0.5mm]
      \multicolumn{1}{c}{eRASS} &
      \multicolumn{1}{c}{Obs.Date} &
      \multicolumn{1}{c}{Offset} &
      \multicolumn{1}{c}{Src.Cnts} &
      \multicolumn{1}{c}{Det.ML} &
      \multicolumn{1}{c}{Rate} \\
      
      \multicolumn{1}{c}{} &
      \multicolumn{1}{c}{} &
      \multicolumn{1}{c}{[arcsec]} &
      \multicolumn{1}{c}{[$\times 10^{3}$]} &
      \multicolumn{1}{c}{} &
      \multicolumn{1}{c}{[cnt/s]} \\

      \midrule[0.5mm]
      1 & 2019/12/20$-$27 & 1.80 & 1.12$\pm$0.034 & 5.9e3 & 7.14$\pm$0.21\\ 
      2 & 2020/06/21$-$27 & 2.74 & 0.96$\pm$0.031 & 5.1e3 & 7.68$\pm$0.25\\ 
      3 & 2020/12/22$-$28 & 1.05 & 0.78$\pm$0.028 & 4.0e3 & 6.30$\pm$0.23\\ 
      4 & 2021/06/24$-$30 & 1.11 & 0.72$\pm$0.027 & 3.7e3 & 6.23$\pm$0.24\\ 
      5 & 2021/12/26$-$2022/01/01 & 0.99 & 0.84$\pm$0.029 & 4.3e3 & 6.65$\pm$0.23\\ 
     
      \bottomrule[0.5mm]
      \end{tabular}
      \tablefoot{eROSITA survey (col.1), actual observational period of GJ\,504 (col.2),  offset between proper motion corrected expected position and X-ray source (col.3), net source counts and detection maximum likelihood (cols.4 and 5) and count rate (col.6). The energy band adopted for the extraction is $0.2-5.0$\,keV, see Sect.~\ref{Sec:eROSITA}.}
\end{center}
\end{table*}
\begin{figure*}
\begin{center}
\parbox{18cm}
{
\parbox{6cm}{\includegraphics[width=0.33\textwidth]{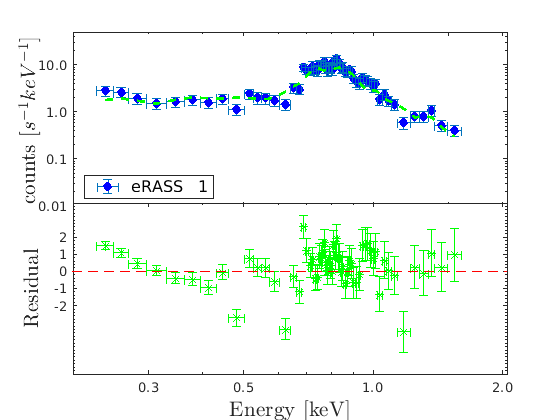}}
\parbox{6cm}{\includegraphics[width=0.33\textwidth]{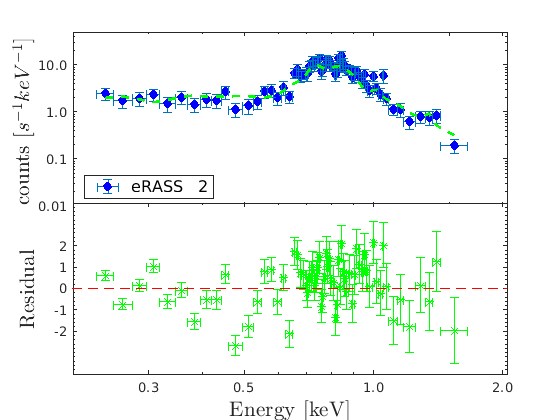}}
\parbox{6cm}{\includegraphics[width=0.33\textwidth]{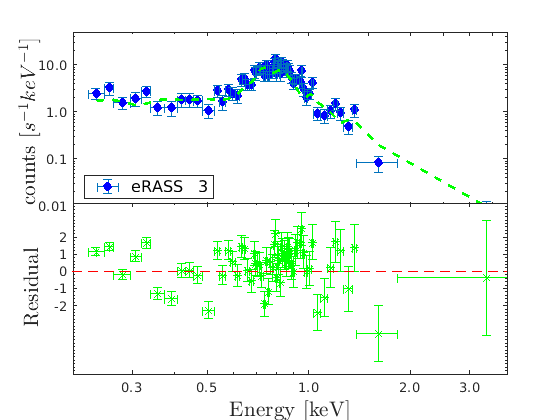}}
}
\parbox{18cm}
{
\parbox{6cm}{\includegraphics[width=0.33\textwidth]{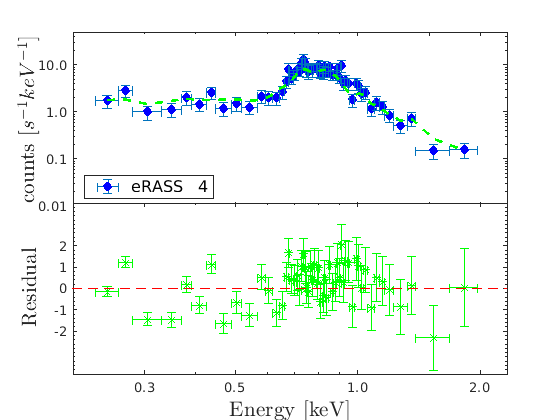}}
\parbox{6cm}{\includegraphics[width=0.33\textwidth]{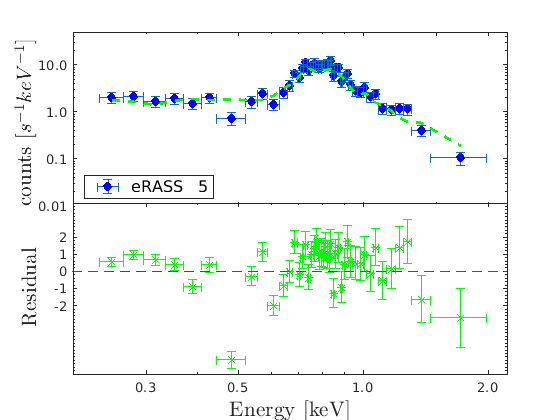}}
\parbox{6cm}{\caption{eROSITA spectra of GJ\,504 plotted together with the best-fitting apec model and the
residuals (both shown in green); see text in Sect.~\ref{subsec:er_spec_anal} and Table~\ref{tab:xspec_par}.}\label{fig:er_spec}}
}

\end{center}
\end{figure*}

\subsection{eROSITA spectral analysis}
\label{subsec:er_spec_anal}
We created the spectrum (shown in Fig.~\ref{fig:er_spec}), response matrix and ancillary file for each eROSITA detection using the {\sc srctool} pipeline with the option AUTO for the choice of the source and background region sizes. 
We grouped each spectrum with $10$ photons per channel for all eRASS, but for eRASS\,1~\&~5 for which we adopted $15$ photons per channel. 
We fitted the $4$ spectra with XSPEC version 12.13 \citep{Arnaud96.0} adopting the XSPEC library for the stellar abundances compiled by \cite{Asplund2009}. 
We used a one-temperature thermal APEC model with global coronal abundance (Z) fixed at $0.3\,Z_{\odot}$, as it is typically considered for stellar coronae \citep{Maggio07.0}.
We calculated the model-dependent X-ray fluxes with the dedicated XSPEC pipeline {\sc flux} initially adopting the eROSITA energy band. For a consistent comparison with the X-ray luminosity from ROSAT  \citep{Voges1999}, we performed a calculation also in the ROSAT energy band ($0.1-2.4$\,keV), finding that the difference in emitted flux within the two bands is not significant, as previously showed by \citet{Magaudda22.0}. We used the model-dependent flux with {\it Gaia}-DR3 distance of $17.58$\,pc to calculate the X-ray luminosity ($\rm L_{x}$). The results of our spectral analysis are summarized in Table~\ref{tab:xspec_par}, where we show the eROSITA survey (col.1), the coronal temperature and emission measure with their $1\,\sigma$ uncertainties (col.2\&3), the reduced chi-squared ($\chi^{2}_{\rm red}$) with the degrees of freedom (cols.4\&5) and the X-ray luminosity in the energy band of $0.1-2.4$\,keV. We notice that X-ray luminosity derived in this work with average value $\rm 10^{29.42}~erg/s$ is in line with the one in \citet{Foster2022} ($\rm 10^{29.4}~erg/s$).

\begin{table}
\begin{center}
 \captionsetup{justification=centering}
\caption{Best-fit parameters from the spectral analysis of GJ\,504 eROSITA spectra.}
    \label{tab:xspec_par}
    \begin{adjustbox}{width=1\linewidth}
    \begin{tabular}{cccccc}
      \midrule[0.5mm]
      \multicolumn{1}{c}{eRASS} &
      \multicolumn{1}{c}{kT} &
      \multicolumn{1}{c}{$\log(EM)$} &
      \multicolumn{1}{c}{$\chi_{\rm red}^{2}$} &
      \multicolumn{1}{c}{d.o.f.} &
      \multicolumn{1}{c}{$\log(L_{\rm x})$} \\
      
      \multicolumn{1}{c}{} &
      \multicolumn{1}{c}{[keV]} &
      \multicolumn{1}{c}{[$cm^{-3}$]} &
      \multicolumn{1}{c}{} &
      \multicolumn{1}{c}{} &
      \multicolumn{1}{c}{[erg/s]}\\

      \midrule[0.5mm]
      1 & 0.64$\pm$0.03 & 51.36$\pm$0.02 & 1.3 & 59 & 29.45$\pm$0.01 \\ 
      2 & 0.61$\pm$0.04 & 51.38$\pm$0.02 & 1.2 & 66 & 29.47$\pm$0.01 \\ 
      3 & 0.56$\pm$0.04 & 51.30$\pm$0.02 & 1.6 & 60 & 29.37$\pm$0.01\\ 
      4 & 0.61$\pm$0.04 & 51.30$\pm$0.02 & 0.9 & 54 & 29.39$\pm$0.01\\ 
      5 & 0.61$\pm$0.04 & 51.31$\pm$0.02 & 1.9 & 43 & 29.40$\pm$0.01\\ 
     
      \bottomrule[0.5mm]
      \end{tabular}
      \end{adjustbox}
      \tablefoot{Survey (col.1), coronal temperature and emission measure (cols.2 and 3), reduced chi-squared and degree of freedom (cols.4 and 5) and X-ray luminosity in the energy band of $0.1-2.4$\,keV (col.6). See Sect.~\ref{subsec:er_spec_anal}.}
\end{center}
\end{table}

\section{Compared evolutionary scenarios: $\rm \Omega_{surf}$ and $\rm L_X$}
\label{Sec:SPI}

Once the best-fit model of GJ 504 is computed as described in Sec.~\ref{Sec:Opt}, then it is coupled to our SPI code in order to compute the evolution of the host-star surface rotation rate ($\rm \Omega_{surf}$) and X-ray luminosity ($\rm L_{x}$) in two different evolutionary scenarios: on one hand, we simulate the evolution of the host star by assuming that there is no close-by, massive planet (``Star without close-by planet'', see Sect.~\ref{Sub: Single Star}); on the other hand, we assume that a massive planet formed in a short orbit ($\rm a_{in} \leq 0.1~AU$)  (or migrated close to its host star before the dissipation of the protoplanetary disc), and study the impact of its eventual inward migration driven by tidal dissipation and engulfment by the host star on the stellar properties (``Star with close-by planet'', see Sect.~\ref{Sub: close-by planet}).

With the rotational history of GJ 504 being unknown, we compute models representative of different rotators, from the super-slow to fast one, by using different values for the initial surface rotation rate ($\rm \Omega_{in} = 1, 2, 3.2, 5, 6, 7, 8, 9, 10, 18~\Omega_{\odot}$). For what concerns the lifetime of the protoplanetary disc, we considered $\rm \tau_{\rm dl} = 2$ Myr for $\rm \Omega_{in} = 18~\Omega_{\odot}$, and $\rm \tau_{\rm dl} = 6$ Myr for the other values. We recall that the choice of the $\rm \Omega_{in}$ and $\rm \tau_{dl}$ is determined from the distribution of surface rotation rates observed for stars in star-forming regions and young open clusters at various ages \citep{Eggenberger2019a}\footnote{Typically in \citet{Eggenberger2019a} the values $\rm \Omega_{in} = 3.2, 5, 18~\Omega_{\odot}$ correspond to the slow, moderate and fast rotators, respectively.}. 

\subsection{Star without close-by planet'' (Sw/oP)}
\label{Sub: Single Star}

In this evolutionary scenario, we assume that there is no close-by, massive planet orbiting around GJ 504. Even if the detection of a sub-stellar companion through direct imaging was announced by \citet{Kuzuhara2013}, with mass $\rm  1 \lesssim M_{c}/M_{J} \lesssim 25 $, this object is too far away from its host star ($\rm \approx 43.5~AU$) to have any significant impact on the evolution of its surface rotation rate and X-ray luminosity, at least according to the physics included in our SPI code.
In this context, in Fig.~\ref{Fig: Single Star}, we collect the results obtained in terms of evolution of the surface rotation rate, stellar Rossby number, and X-ray luminosity. 

\subsubsection{Sw/oP: Surface rotation rate}

In the top panel of Fig.~\ref{Fig: Single Star} we show the evolution of the surface rotation rate normalised to the value of the Sun ($\rm \Omega_{\odot} = 2.9 \times 10^{-6}~s^{-1}$) for each of the considered rotational histories, as function of the stellar age. As expected in the case of stars without close-by planets, the evolution of the surface rotation rate is initially governed by the contraction of the stellar radius during the PMS phase, until when the star reaches the Zero Age Main Sequence (ZAMS), at about $\rm 25~Myr$. Subsequently, the braking of the stellar surface due to magnetised winds takes over, determining the overlap of the majority of the tracks at $\rm \approx 500~Myr$, except for the $\rm 1~\Omega_{\odot}$ model, that globally evolves at lower values. By comparing the theoretical tracks with the rotation rate derived for GJ 504 in \citet{DiMauro2022} ($\rm P_{rot} = 3.4$ days, $\rm \Omega_{GJ 504} = 2.13\times 10^{-5}~s^{-1}$), we notice that there is compatibility only if we consider the youngest possible value for the stellar age ($\rm 0.36$ Gyr), and only for $\rm \Omega_{in} \geq 10~\Omega_{\odot}$, otherwise $\rm \Omega_{GJ 504}$ is larger than what expected from our models.

We analyse whether considering the onset of weakened magnetic braking at $\rm Ro = Ro_{cri}$ could result in a better overlap of the theoretical tracks with the observational data. Because of the impact of this mechanism indeed, the braking of the stellar surface would be stalled after a certain age, leading to larger surface rotation rates at the age of GJ 504. The trend of the evolutionary tracks in this circumstance, computed only for $\rm \Omega_{in} = 3.2, 5, 18~\Omega_{\odot}$, is showed by the grey-dashed lines in the top panel of Fig.~\ref{Fig: Single Star}. It is possible to noticed that at the onset of weakened magnetic braking ($\rm \sim 2~Gyr$), the star has already been significantly braked, and there is no compatibility with the observational value.

\subsubsection{Sw/oP: Stellar Rossby number}

\begin{figure}[h!]
\includegraphics[width=\linewidth]{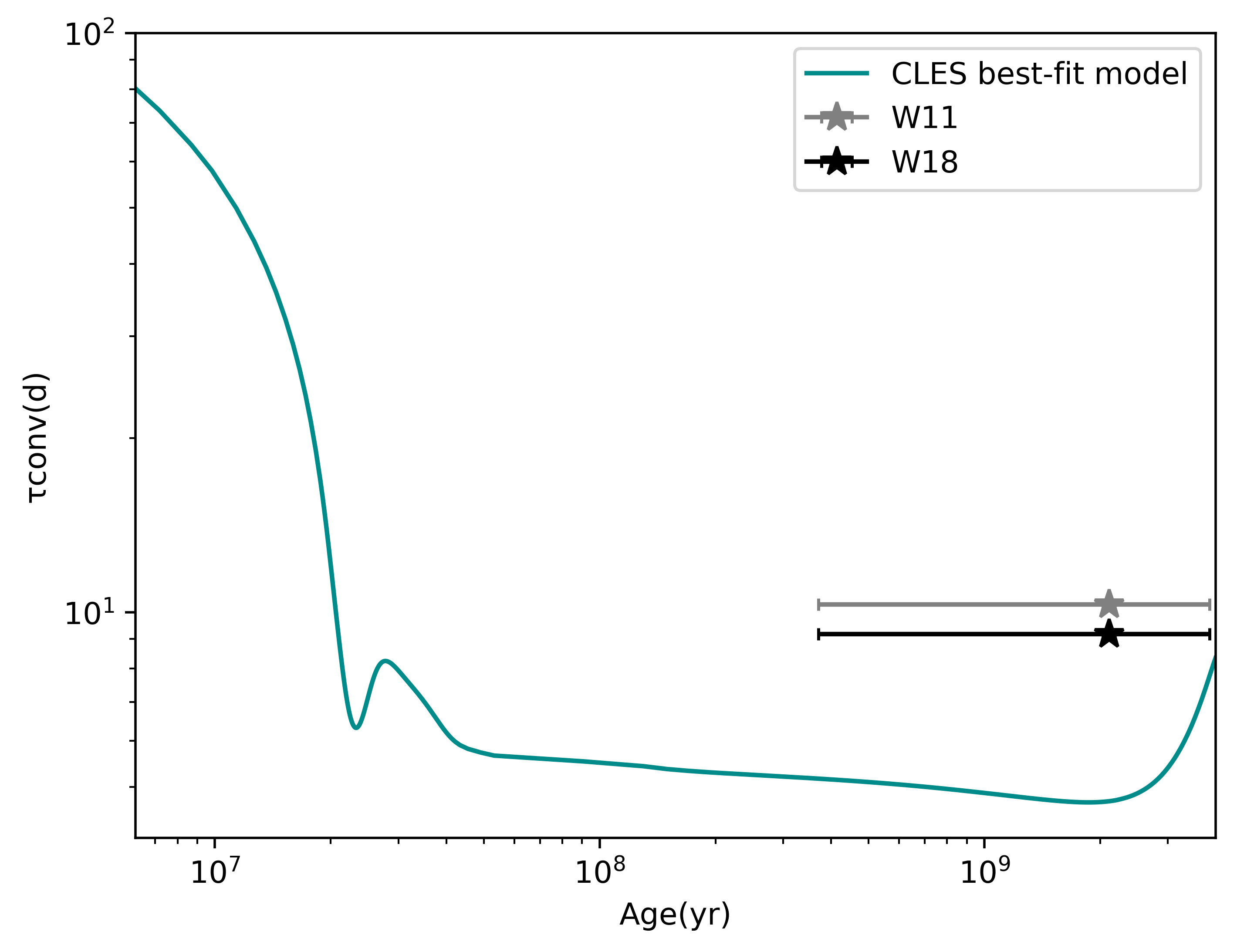}
\caption{Comparison between the convective turnover timescale obtained from CLES best-fit model by using Eq.~\ref{Eq:tauconv}, and the values computed as in \citet{Wright2011} (gray star) and \citet{Wright2018} (black star).}
\label{Fig: tconv}
\end{figure}

In the middle panel of Fig.~\ref{Fig: Single Star}, the stellar Rossby number evolutionary tracks are presented. To make a comparison with the Rossby number obtained by using $\rm P_{rot} = 3.4~d$, and given the significant uncertainty on GJ 504 age, we estimate three different $\rm R_{O}$, at the lower, mean and upper limit of the errorbar. By using Eq.~\ref{Eq:tauconv}, we derive: $\rm R_{O_{low}} = 0.65$, $\rm R_{O_{m}}= 0.62$, $\rm R_{O_{up}} = 0.43$. Similarly to what we found for the surface rotation rate, an agreement with the evolutionary tracks is obtained only for the smallest value of the host star age's, namely for $\rm R_{O_{low}}$, and for $\rm \Omega_{in} \geq 10~\Omega_{\odot}$. It is worth noticing that we use a model-dependent approach to compute the Rossby number by means of Eq.~\ref{Eq:tauconv} for $\rm \tau_{conv}$, to be consistent with the physics included in the SPI code. However, it is possible to compare our estimation of $\rm \tau_{conv}$ with the values derived by prescriptions based on purely observational properties. This is the case of the formulae in \citet{Wright2011,Wright2018} based on the $(V - K_{S})$ colour, which were calibrated on a sample of solar, and late type stars on the MS. In order to make a comparison with our model-dependent convective turnover timescale computation, we thus use Eq. 10 in \citet{Wright2011} and Eq. 5 in \citet{Wright2018}, with $ (V - K_{S}) = 1.29$ \citep{Wright2011}. We finally obtain: $\rm \tau_{conv_{W11}} = 10.3228~d$, and $\rm \tau_{conv_{W18}} = 9.1727~d$. In Fig.~\ref{Fig: tconv} we compare these two values with the evolutionary track. At $\rm \approx 2.11~Gyr$, there is a difference of about $\rm 8~d$ between the purely observational and theoretical estimations of $\rm \tau_{conv}$. Consequently, also the corresponding Rossby numbers show significantly diverse values in the middle panel of Fig.~\ref{Fig: Single Star}, where $\rm R_{O_{m}}= 0.62$, $\rm R_{O_{W11}}= 0.33$ and $\rm R_{O_{W18}}= 0.37$.

In the same panel, we also indicate the value of the critical Rossby number ($\rm R_{O_{cri}}$, black-dotted line, \citet{Metcalfe2024}), above which the star may transition to the weakened magnetic braking regime \citep{VanSaders2016}. According to our theoretical models, a star like GJ 504 would cross $\rm R_{O_{cri}}$ at an age of $\rm \sim 1-2$ Gyr. As mentioned in the previous sections, at this point the star has been significantly braked by the magnetised winds, to values well below $\rm \Omega_{GJ504}$. Thus, magnetic braking seems to not be responsible for the faster rotation of the star.

\subsubsection{Sw/oP: X-ray luminosity}

\begin{figure}[h!]
\includegraphics[width=\linewidth]{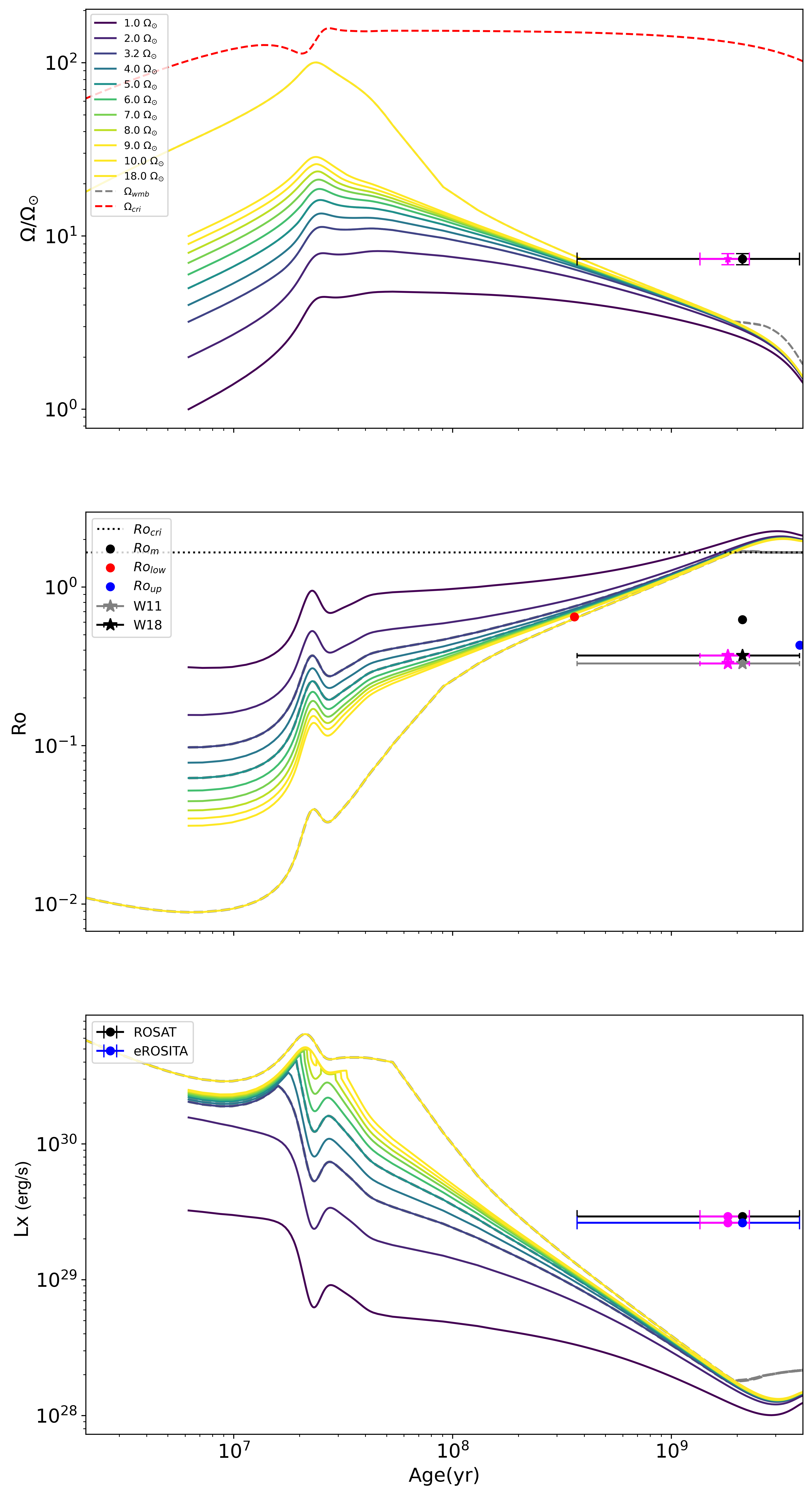}
\caption{\small{\emph{Top panel:} Surface rotation rate's evolution vs age for our optimal stellar model, with $\rm \Omega_{in} = 1, 2, 3.2, 4, 5, 6, 7, 8, 9, 10, 18~\Omega_{\odot}$. The red-dashed line shows the evolution of the critical rotation velocity ($\Omega_{\rm crit}$), defined as the velocity at which the centrifugal acceleration at the equator equals the gravity. The magenta/black markers show GJ 504's surface rotation rate, with age uncertainties derived from the global and local minimisation modelling, respectively. \emph{Middle panel:} Evolution of the stellar Rossby number ($\rm R_{O}$). $\rm R_{O}$ for GJ 504 is indicated by the red, black, and blue circles corresponding to the lower, mean and upper values of the largest age uncertainty, respectively. The gray and black stars represent the values obtained from $\rm \tau_{conv}$ as in \citet{Wright2011} and \citet{Wright2018}. Analogously, the magenta markers represent the same quantities for the smallest age uncertainty. \emph{Bottom panel:} Evolution of the X-ray luminosity for each of the considered rotators. The black and blue markers show the X-ray luminosities from \citet{Voges1999, Wright2011} (ROSAT) and in this work (eROSITA), respectively. Analogously, the magenta markers represent the same quantities for the smallest age uncertainty. In all panels the starting point of the tracks corresponds to the dissipation of the protoplanetary disc (2 Myr for $\Omega_{in} = 18~\Omega_{\odot}$, 6 Myr for the other ones).}}
\label{Fig: Single Star}
\end{figure}

In the bottom panel of Fig.~\ref{Fig: Single Star}, the X-ray luminosity tracks are showed. Each track corresponds to a different rotational history. We compare the tracks with observational data from \citet{Voges1999, Wright2011} for ROSAT, with $\rm L_{X} = 10^{29.47}~erg/s$ (black dot), and from our analysis of eROSITA, from which we derive an average value $\rm L_{X} = 10^{29.42}~erg/s$. Differently from what we have observed for $\rm \Omega$ and $\rm Ro$, in this case, an agreement between the theoretical tracks and the observed value is not observed, even for the lowest age value allowed for GJ 504. It is worth recalling that low and solar-like stars may experience significant variations in the emission of X-ray luminosity, over one order of magnitude, usually linked with the periodicity of their magnetic cycle.

\subsection{Star with close-by planet (SwP)}
\label{Sub: close-by planet}

In the star with close-by planet scenario, we test the impact of the inward migration of a giant planet on the surface rotation rate and X-ray luminosity of the host star, GJ 504. On the base of analytical considerations, \cite{Fuhrmann2015} proposed that the minimum mass for the planet to reproduce the observed stellar spin up is $\rm M_{pl} \gtrsim 2.7~M_{J}$. In their work, the authors assumed that the angular momentum of the star is entirely related with the transfer of angular momentum from the planetary orbit, that occurred once the planet filled its Roche lobe. In this work, we account for the intrinsic angular momentum of the host star, together with the one transferred by the planetary orbit. Thus we study the impact of equilibrium/dynamical tides dissipated within the host star convective envelope on the migration of the planet, and its feedback on the stellar surface rotation rate and X-ray luminosity evolution. 

\begin{figure}
\centering
\subfigure{\includegraphics[width=\linewidth]{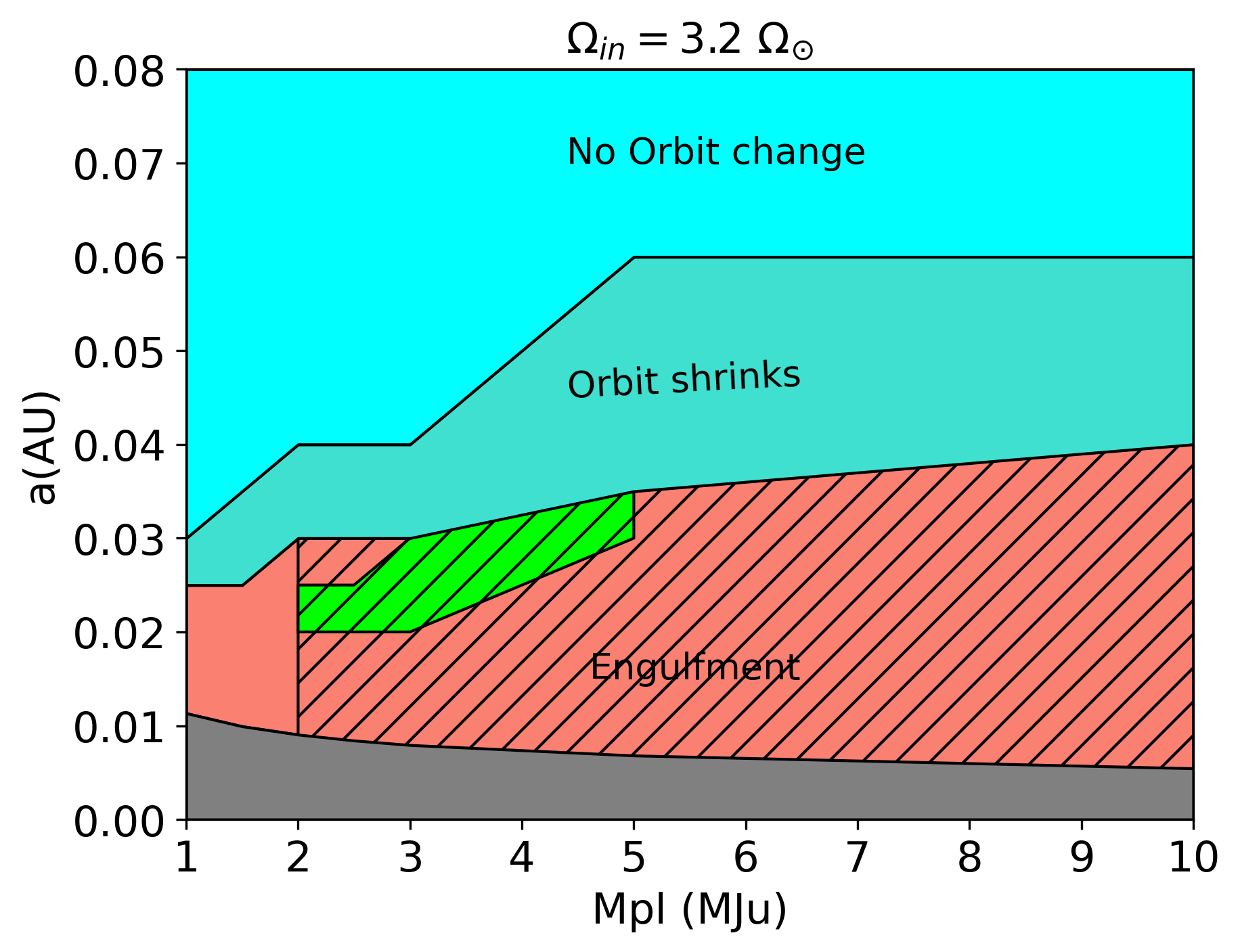}}
\subfigure{\includegraphics[width=\linewidth]{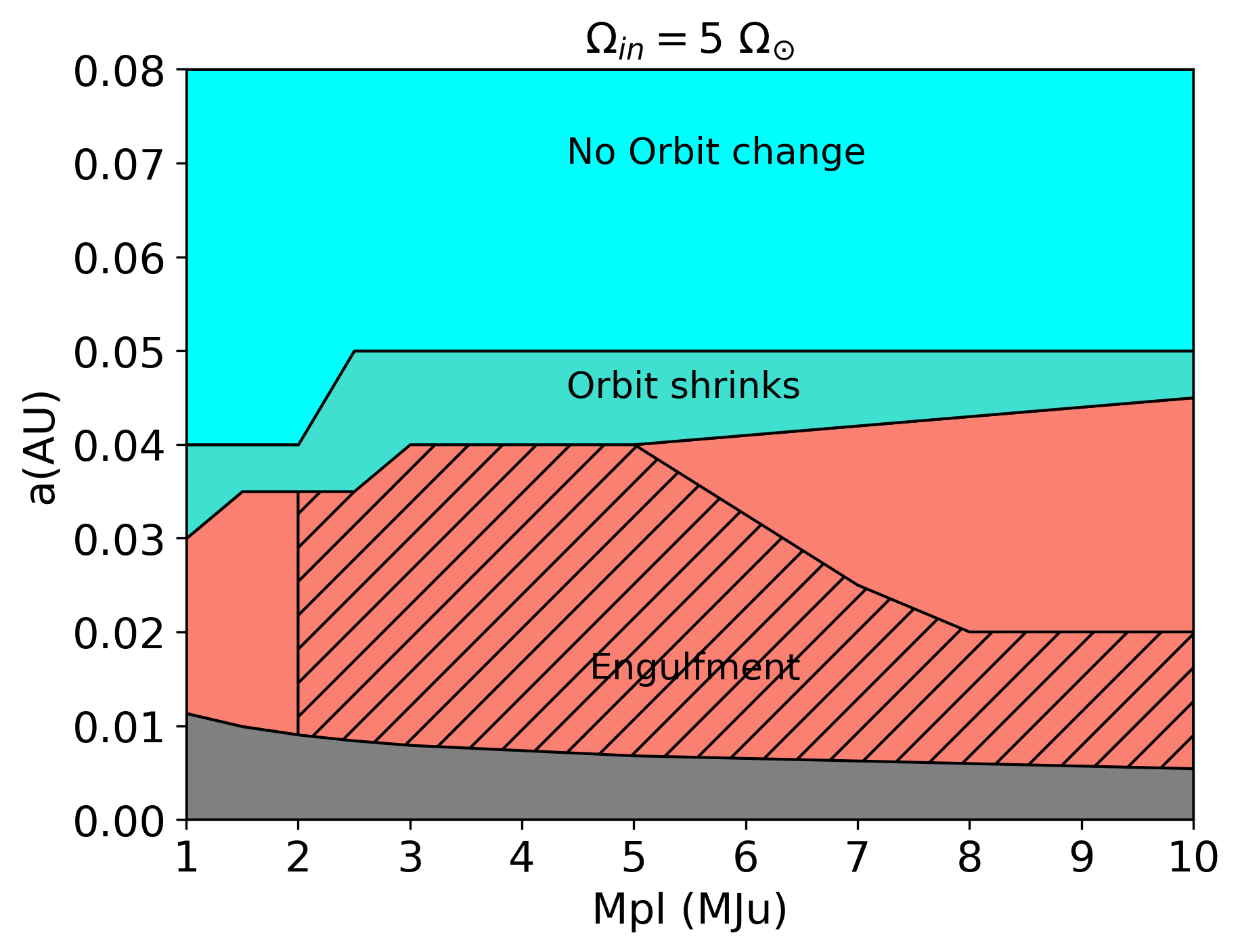}}
\subfigure{\includegraphics[width=\linewidth]{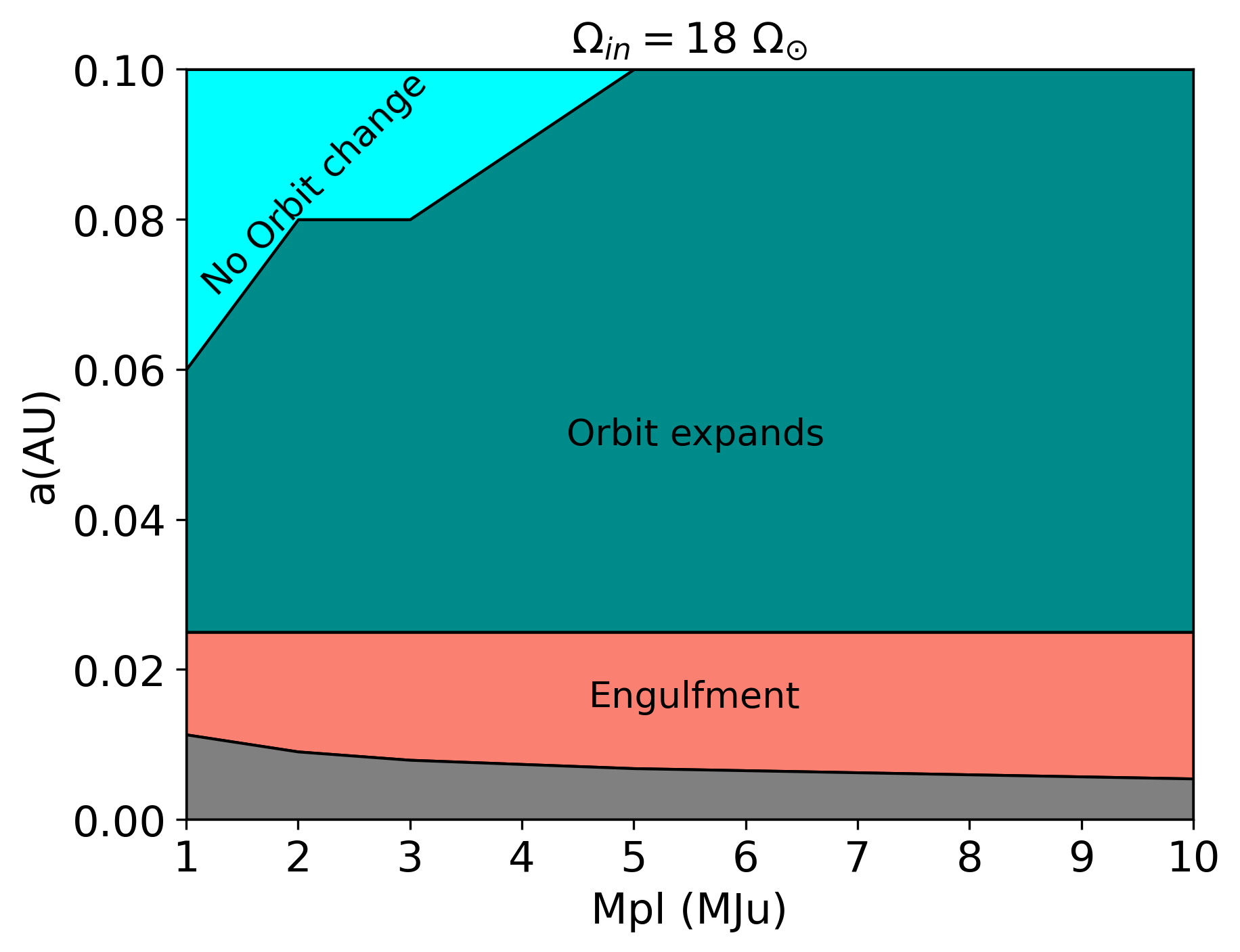}}
\caption{Different orbital evolutions for a putative planet with initial planetary mass $\rm 1 \leq M_{pl}/M_{J} \leq 10$ and initial orbital distance $\rm 0.02 \leq a~(AU) \leq 0.1$, computed for $\rm \Omega_{in} = 3.2~\Omega_{\odot}$ (Top panel),  $\rm \Omega_{in} = 5~\Omega_{\odot}$ (Middle panel), and $\rm \Omega_{in} = 18~\Omega_{\odot}$ (Bottom panel). The salmon-hatched area indicates the region in $\rm M_{pl}$ and $\rm a$ for which, after the engulfment, it is possible to reproduce the rotational period and X-ray luminosity of GJ 504 when considering the largest age uncertainty. The lime-hatched area indicates a subset of the salmon-hatched area, respect to the smaller age uncertainty. The turquoise area in the Top and Middle panels shows the region in which the planetary orbit shrinks, while the dark-cyan area in the Bottom panel indicates where the planetary orbit expands. The gray shaded area shows the region below the Roche limit.}
\label{Fig:Param}
\end{figure}

\begin{figure}
\centering
\subfigure{\includegraphics[width=\linewidth]{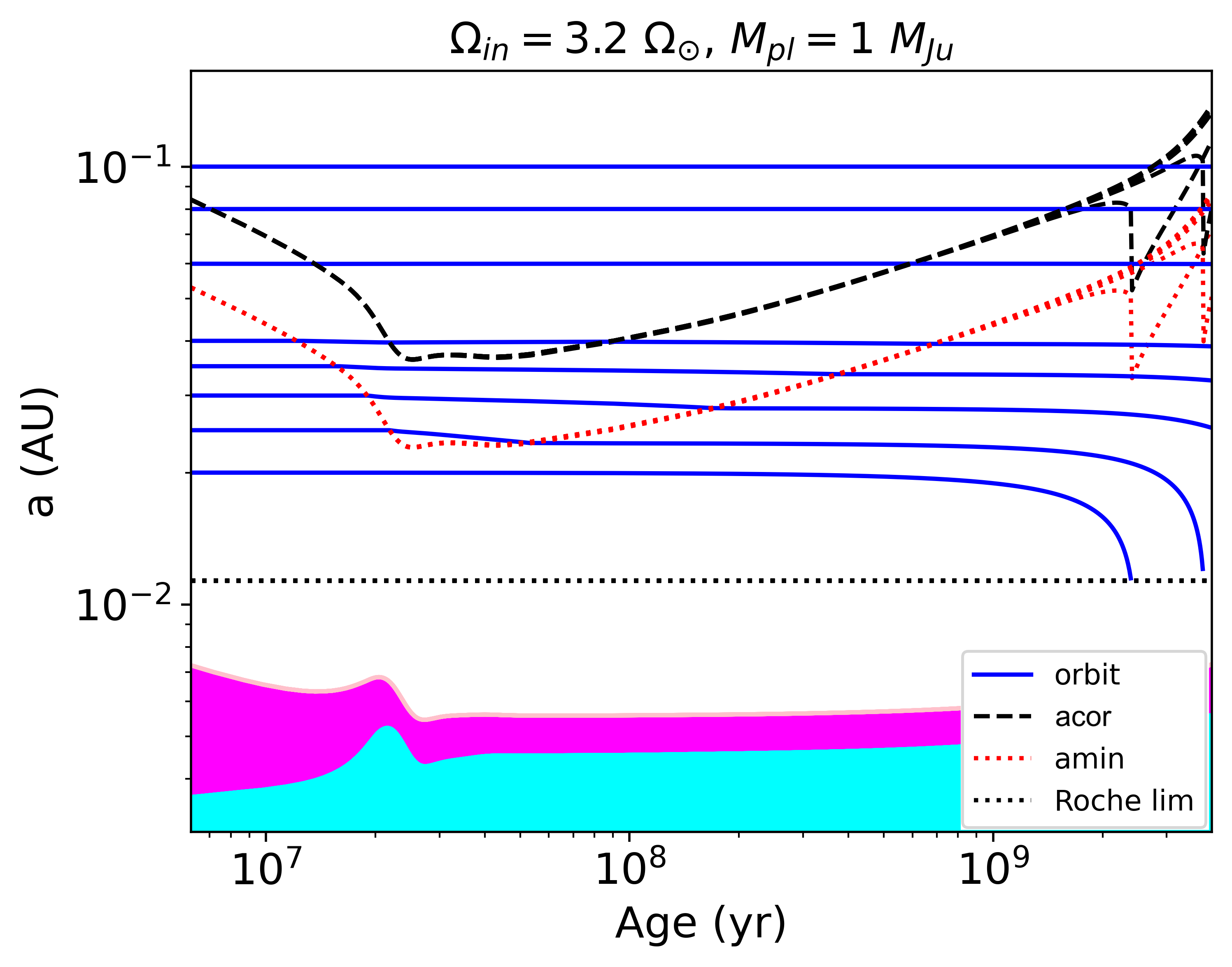}}
\subfigure{\includegraphics[width=\linewidth]{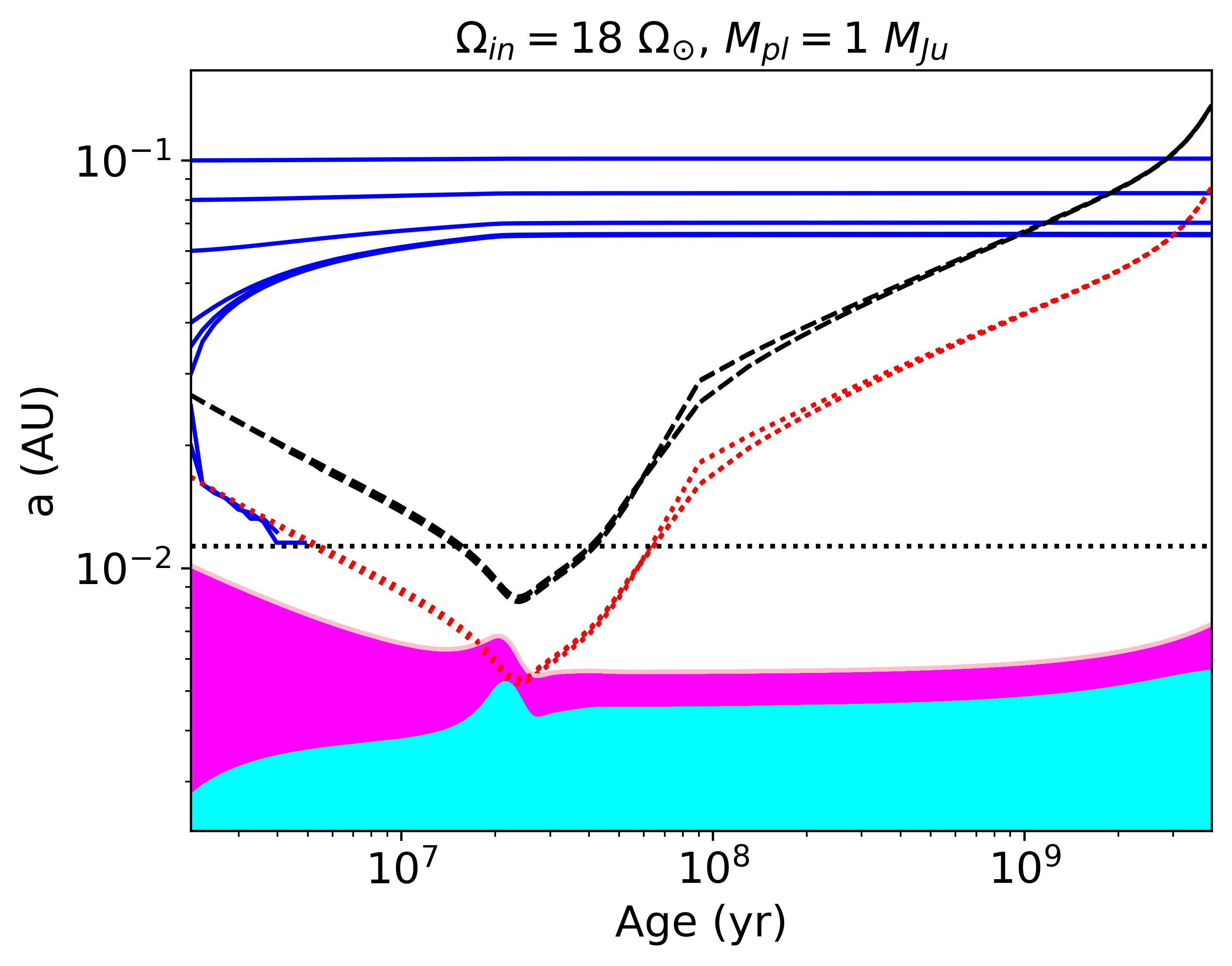}}
\caption{Orbital evolution for a $\rm 1~M_{J}$ planet with $\rm 0.02 \leq a_{in}~(AU) \leq 0.1$, and $\rm \Omega_{in} = 3.2, 18~\Omega_{\odot}$ (Top and Bottom panels, respectively). The blue solid lines show the evolution of the orbital distance. The red-dotted and black-dashed lines indicate the evolution of $ a_{\rm min}$ and $\rm a_{cor}$, respectively. The Roche limit is showed by the black dotted line. The cyan and magenta shaded areas represent the extension of the stellar core and envelope along the evolution of the star, respectively. }
\label{Fig:MultipleOrbits}
\end{figure}

At first, we explore a range of initial orbital distances ($\rm 0.02 \leq a~(AU) \leq 0.1$), planetary masses ($\rm  1 \leq M_{pl}/M_{J} \leq 10$), and select stellar models representative of a slow, moderate and fast rotators, namely $\rm \Omega_{in} = 3.2, 5, 18~\Omega_{\odot}$ \citep{Eggenberger2019a}, to find for which combination the computed stellar rotational period and X-ray luminosity are compatible with the observed values. In Fig.~\ref{Fig:Param} we show the corresponding results. From these computations, we find that for the slow and moderate rotators it is possible to determine a region of compatibility in the $\rm a-M_{pl}$ plane, indicacted by the black-hatched area in the top and middle panels of Fig.~\ref{Fig:Param}. Contrarily, for the fast rotator case this compatibility is not retrieved. In this case the planetary engulfment generally occurs at too early evolutionary stages ($\rm \sim 4~Myr$), when the star is spinning up due to its contraction on the PMS, and thus the contribution to the stellar angular momentum is not only less important, but also quickly erased by subsequent magnetic braking. A further difference between the slow-moderate and fast rotator is that while in the first two cases the orbit of the planet shrinks for $\rm 0.03 \lesssim a(AU) \lesssim 0.06$, in the last one the orbit expands for $\rm 0.025 \lesssim a(AU) \lesssim 0.1$. This is due to the different geometry of the corotation radius ($\rm a_{cor}$), and consequently $\rm a_{min}$, that at the beginning of the evolution is smaller for larger values of $\rm \Omega_{in}$ (see Eq.~\ref{Eq:coradius}). To clarify this point, in Fig.~\ref{Fig:MultipleOrbits} we show an example of orbital evolution for a $\rm 1~M_{J}$ planet with $\rm a_{in} = 0.02, 0.025, 0.03, 0.035, 0.04, 0.06, 0.08, 0.1~AU$, and $\rm \Omega_{in} = 3.2, 18~\Omega_{\odot}$. In the slow rotator case, most of the tracks ($\rm a_{in} \leq 0.06~AU$) begin the evolution below $\rm a_{min}$, thus dynamical tides are not at work and the orbit of the planet remains constant. Even after crossing $\rm a_{min}$ at about $\rm 10-20~Myr$, the orbit is not significantly affected by dynamical tides, given that the initial surface rotation rate of the star is quite small, and the efficiency of tidal dissipation is tightly linked with this quantity. Equilibrium tides finally deflect the orbit for $\rm a_{in} = 0.02, 0.025~AU$ at $\rm \sim 2~Gyr$, making the planet migrate towards its host star, since $\rm a_{in} < a_{cor} $. It is worth noting that we follow the evolution of the planet till when it reaches its Roche limit\footnote{$\rm Roche_{lim} \approx 2R_{pl} (M_{\star}/M_{pl} )^{1/3} $} \citep{ZhangPenev2014}. Following \citet{Benbakoura2019}, at this point we assume that it gets instantaneously depleted by tidal interactions and transfers its orbital angular momentum to the host star. As indicated in \citet{Metzger2012}, when the ratio between the planetary and stellar density is in the range $\rm 1 < \rho_{pl}/\rho_{\star} < 5$, and the planet overflows its Roche lobe, an unstable mass transfer takes place tearing apart the planet within a timescale of several hours. For the range of planetary masses considered here, for the slow and moderate rotators, the ratio $\rm \rho_{pl}/\rho_{\star}$ at the engulfment ($\rm \sim 2~Gyr$) is about 2-3. Thus the assumption of instantaneous planetary disruption seems to be reasonable. For what concerns the fast rotator case, as it is shown in Fig.~\ref{Fig:MultipleOrbits}, dynamical tides are much more efficiently dissipated and planets starting the evolution below the corotation radius quickly migrate towards the host star, reaching the Roche limit. At this evolutionary stage, the ratio $\rm \rho_{pl}/\rho_{\star} >5$, and according to \citet{Metzger2012} the planet would spiral and finally plunge into the stellar atmosphere. This event probably lasts more than the aforementioned instantaneous depletion. Even if we assume an instantaneous angular momentum transfer, to maximise its effect, given that the star is still spinning up at this stage, as mentioned before it does not significantly impact the rotational evolution of its surface (at least according to the physics included in this work). Planets starting their evolution above the corotation radius ($\rm 0.025 <a_{in}~(AU) < 0.08$) are instead pushed outwards.  

Contextually to the orbital evolution, we compute the mass loss from the planet due to the stellar XUV flux. Given the large planetary masses considered in these simulations, the mass loss due to the stellar high energy irradiation turns out to be negligible \citep[e.g.][]{Owen2013}.

After exploring the parameter space in $\rm M_{pl}$, $\rm a_{in}$ and $\rm\Omega_{in}$, in the following we focus on studying the orbital evolution of a planet with mass $\rm 3~M_{J}$, arbitrarily chosen in the range of compatibility, and with $\rm a_{in} = 0.025~AU$, for which we analyse a finer grid in $\rm \Omega_{in} =  1, 2, 3.2, 5, 6, 7, 8, 9, 10, 18~\Omega_{\odot}$, as we did for the ``Star without close-by planet'' scenario case. The goal is to investigate for which range of rotational histories the simulations reproduce the observed $\rm P_{rot}$ and $\rm L_{X}$. The corresponding results are showed in Fig.~\ref{Fig: Star companion}\footnote{For the sake of completeness, in Appendix~\ref{App: StarComp1MJu} we present similar results obtained this time for a $\rm 1~M_{J}$ planet with $\rm a_{in} = 0.02~AU$, showing that there is compatibility between the theoretical and observational values for the rotational period, but not for the X-ray luminosity.}.

It is worth recalling that the results found in this section rely on the assumption that the planet formed, or migrated before the dispersal of the planetary disk, very closely to the host star, sitting on a coplanar-circular orbit. However, other mechanisms could have brought the planet to a very short orbit after its formation, with an initial planetary architecture characterised by a certain inclination and eccentricity. In particular, the presence of sub-stellar companion (at projected distance $\rm \sim 43.5$ AU), could have triggered the inner planet's orbital shrinking by means of plant-planet scattering \citep[e.g.][]{RasioFord1996}, and/or Kozai-Lidov oscillations \citep{Kozai1962, Lidov1962, Mazeh1997}. This last one consists in a dynamical mechanism affecting the orbit of the inner planet with a periodically exchanging variation between its eccentricity and inclination. The excitation of the planetary eccentricity by means of these processes, together with tidal dissipation within the inner planet at every passage at the periastron, would potentially relocate the planet on a much shorter and circular orbit, eventually short enough to induce tidal dissipation in the stellar host. In a more extreme scenario, for relatively large eccentricity values, the planet could get directly engulfed at one of its periastron passages.
Depending on the properties of the perturber, the initial orbital parameters of the system, and the structure of the inner planet, the timescales for these mechanisms to be triggered and result in an efficient planetary engulfment and spin-up of the host star might significantly vary. In the case of GJ 504, given the limited constrains available on the system properties, it is challenging to establish whether planet-planet scattering and/or Lidov-Kozai oscillations could be triggered. Since stellar tides could still play a major role in a scenario considering also these two processes, future simulations accounting for this mechanism could benefit from the results found in this work.

\subsubsection{SwP: Surface rotation rate}

In the top panel of Fig.~\ref{Fig: Star companion} we show the evolution of the stellar surface rotation rate for each of the considered rotational histories, as function of the stellar age. As for the ``Star without close-by planet'' scenario, the PMS phase is dominated by the structural contraction, until the star reaches the ZAMS and the magnetic braking takes over. Since the planetary companion orbits very closely to the host star, because of the impact of equilibrium/dynamical tides, it migrates inwards. The majority of the orbital angular momentum is transferred to the star when the planet reaches the closest orbits to the Roche limit, and thus the star experiences a kick which results in a spike in the $\rm \Omega$ vs $\rm Age$ track. For $\rm \Omega_{in}  \geq 3.2~\Omega_{\odot}$, the larger the value of $\rm \Omega_{in}$ and the earlier the engulfment of the planet occurs. At the contrary, for $\rm \Omega_{in} < 3.2~\Omega_{\odot}$, since the orbital migration is only driven by equilibrium tides, the engulfment occurs approximately at the same age ($\rm \sim 2.5~Gyr$). As mentioned above, for the fastest rotator considered here ($\rm 18~\Omega_{\odot}$), the planet is engulfed during the PMS and the angular momentum transferred from the planetary orbit does not contribute significantly to the global one of the star, as it is shown from the yellow track in the top panel of Fig.~\ref{Fig: Star companion}, where at the moment of the engulfment ($\rm \sim 10~Myr$) only a small increase of $\rm \Omega$ is visible. For the other rotators instead, the planetary engulfment induces a stellar spin-up to about $\rm 11-12~\Omega_{\odot}$. The quasi-vertical increase of the surface rotation rate is followed by a smoother decay, driven by the magnetic braking. Given that the timescale of the rotation rate decay is larger than the spin-up one, if a planetary engulfment caused the acceleration of GJ 504, it is more probable that we are observing the star in the decelerating phase. In the less likely hypothesis that the planet has not been engulfed yet, but it is at the edge of its Roche limit, transferring its angular momentum to the star, we may try to estimate the radial velocity (RV) semi-amplitude signal induced on its host. By using Eq.~1 in \citet{Cumming1999}, considering the lowest initial mass for the companion compatible to reproduce the observed $\rm P_{rot}$ and $\rm L_{X}$ for GJ 504, namely $\rm 2~M_{J}$, the line-of-sight inclination from \citet{Bonnefoy2018} ($\rm i=18.6$ degrees), and an orbital distance $\rm a_{in} = 0.01~AU$ ($\rm  P_{orb} = 8~d$), we get $\rm K = 52.18~m/s$. \citet{DOrazi2017} rule out the presence of a massive companion, close enough to tidally spin up the central star on the base of RV monitoring performed at Lick Observatory \citep{Fischer2014}. The observed RV dispersion in this case is $\rm 25.7~m/s$, which may be comparable with the activity level of the star. They thus exclude the presence of close-by companions more massive than $\rm 0.5-1~M_{J}$. Nevertheless, if the planet is undergoing tidal destruction when crossing the Roche limit, the actual RV semi-amplitude could be much smaller than the $\rm \sim 52~m/s$ estimated above, since it would be caused only by some planetary remnants at this stage. As reported in \citet{DOrazi2017}, this would require purposely designed observations.

In terms of compatibility between the observational value for the surface rotation rate and the evolutionary tracks, if we look at the decaying branches for each $\rm \Omega_{in}$, it is possible to narrow down the initial surface rotation rate to values $\rm \leq 10~\Omega_{\odot}$. It is worth noting that the overlap between the evolutionary tracks and the observational $\rm \Omega_{\star}$ would favour ages between $\rm 360~Myr$ and $\rm 3~Gyr$, approximately, while no overlap is obtained for older ages.

\subsubsection{SwP: Stellar Rossby number}

Once the star gets spun-up by the planetary engulfment, consequently its Rossby number decreases. This is presented in the middle panel of Fig.~\ref{Fig: Star companion}, where for each of the considered rotators the Rossby number sharply decreases when the planet reaches its Roche limit. In this context, the tracks predicts smaller values of $\rm R_{O}$, which become compatible with both $\rm R_{O_{m}}$ and $\rm R_{O_{up}}$, even though the maximum age for compatibility is $\rm \sim 3~Gyr$, as for the surface rotation rate. A better compatibility is observed also with the Rossby number computed by means of \citet{Wright2011, Wright2018} formulae. 


\subsubsection{SwP: X-ray luminosity}

\begin{figure}
\includegraphics[width=\linewidth]{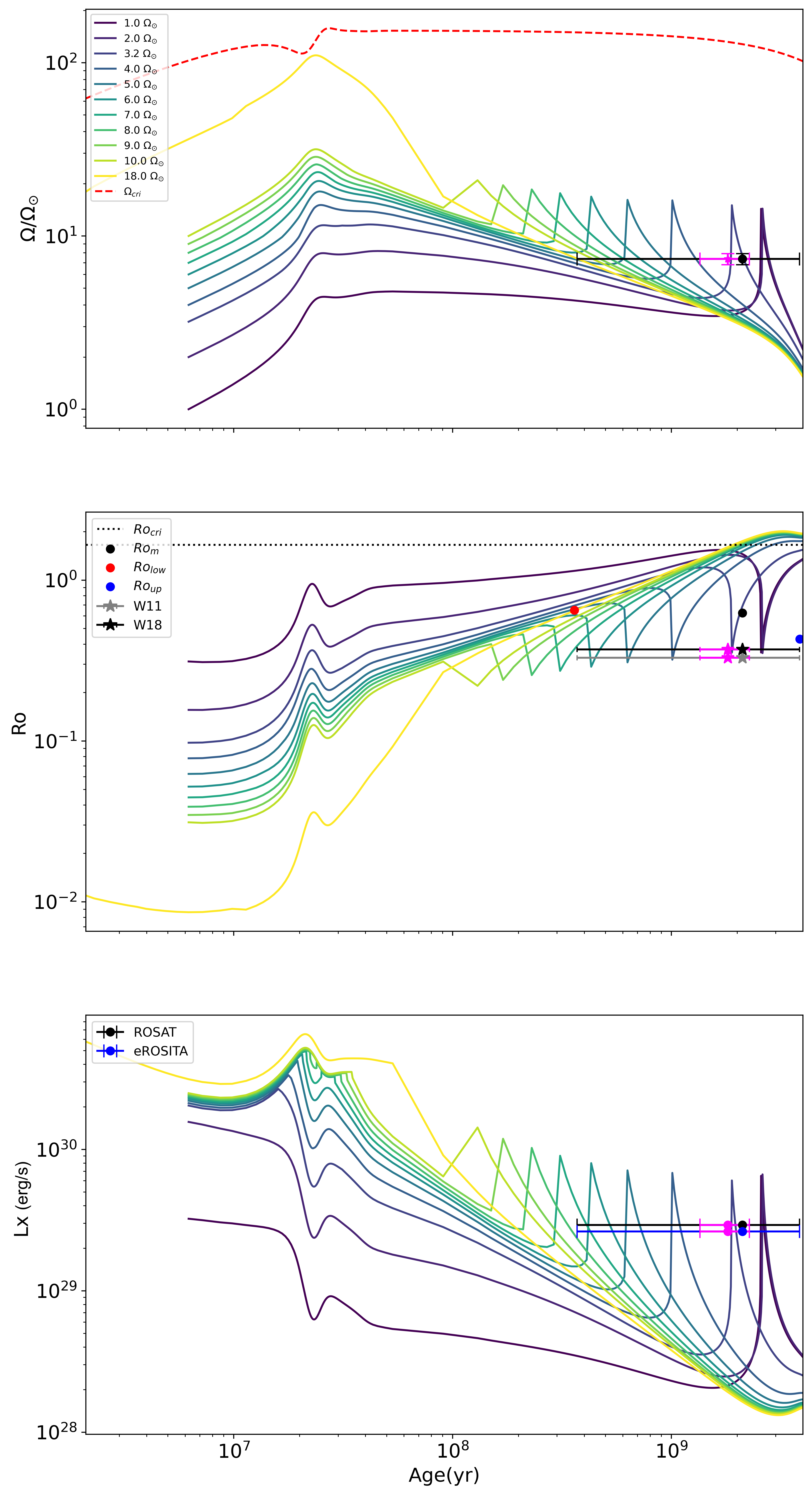}
\caption{\small{Evolution of the stellar surface rotation rate (\emph{Top panel}), stellar Rossby number (\emph{Middle panel}) and X-ray luminosity (\emph{Bottom panel}), under the impact of planetary inward migration. For the stellar initial surface rotation rate we considered the following values $\rm \Omega_{in} = 1, 2, 3.2, 4, 5, 6, 7, 8, 9, 10, 18~\Omega_{\odot}$, for the planetary mass $\rm M_{pl} = 3~M_{J}$, and initial planetary orbital distance $\rm a_{in} = 0.025~AU$. The meaning of the other quantities is the same as in Fig.~\ref{Fig: Single Star}.}}
\label{Fig: Star companion}
\end{figure}

The final comparison between the X-ray luminosity tracks and the observational data obtained from ROSAT \citep{Voges1999, Wright2011} and eROSITA are showed in the bottom panel of Fig.~\ref{Fig: Star companion}. Very interestingly, with the engulfment of a massive enough planet ($\rm M_{pl} \geq2~M_{J}$) and for $\rm \Omega_{in} \lesssim 7~\Omega_{\odot}$, an overlap between the tracks and the observational data is retrieved. Conversely to the comparison with the surface rotation rate, the interval for the possible values of $\rm \Omega_{in}$ compatible with $\rm L_{X_{GJ 504}}$ is narrower. Therefore, in order to reproduce at the same time both the rotational period and X-ray luminosity of GJ 504, the initial surface rotation rate of the star should be $\rm \leq 7~\Omega_{\odot}$.

\section{Conclusion}
\label{Sec:Conclusion}

In this work we study the properties of GJ 504 and investigate whether with the engulfment of a planetary companion the observed surface rotation rate and X-ray luminosity can be reproduced.

On the base of the best-fit stellar model of GJ 504 derived by means of a two-steps minimisation technique, and by using our SPI code, we compute tracks for the stellar surface rotation rate and X-ray luminosity assuming two evolutionary scenarios: ``Star without close-by planet'' in which there is no close-by companion influencing the evolution of the star, and ``Star with close-by planet'' in which a massive planet orbits closely and migrates inwards due to tides dissipated within the host star. The rotational history of the star being unknown, we select a broad range of initial surface rotation rates, from super-slow ($\rm 1~\Omega_{\odot}$) to fast rotator ($\rm 18~\Omega_{\odot}$). When comparing the tracks obtained in the ``Star without close-by planet'' evolutionary scenario, we find that there is compatibility with the surface rotation rate of GJ 504 only for $\rm \Omega_{in} \geq 10~\Omega_{\odot}$, within the considered age range $\rm (0.36 - 3.86)~Gyr$. There is no compatibility instead with the observed X-ray luminosity, for which the tracks predict smaller values. 

In the ``Star with close-by planet'' scenario, we initially explore the parameter space in planetary initial mass ($\rm 1 \leq M_{pl}(M_{J}) \leq 10$), orbital distance ($\rm 0.02 \leq a_{in}(AU) \leq 0.1$) and host star surface rotation rate ($\rm \Omega_{in}  = 3.2, 5, 18~\Omega_{\odot}$) to find the set of conditions needed for the planet to efficiently spin-up the star after the engulfment. As a result, we find that for $\rm M_{pl} \geq 2~M_{J}$, $\rm 0.02 \leq a_{in}(AU) \leq  0.035$ and host star $\rm \Omega_{in} < 18~\Omega_{\odot}$, the engulfed planet is able to spin-up the host star to values compatible with both the surface rotation rate and X-ray luminosity observed for GJ 504.

Given the broad range of the considered host star surface rotation rates, within the aforementioned parameter space, we arbitrarily focused on studying the orbital evolution of a $\rm 3~M_{J}$ planet, at $\rm a_{in} = 0.025~AU$, and a finer grid in $\rm \Omega_{in} =  1, 2, 3.2, 5, 6, 7, 8, 9, 10, 18~\Omega_{\odot}$, with the aim of breaking down the degeneracy on the GJ 504 initial surface rotation rate. In this star-planet configuration, we find that the surface rotation rate and X-ray luminosity of GJ 504 can be both reproduced for $\rm \Omega_{in} \leq 7~\Omega_{\odot}$. Moreover, for $\rm 3.2 \leq \Omega_{in}/ \Omega_{\odot}\leq 7$, the larger the initial surface rotation rate, and the earlier the engulfment occurs. An interesting result is that the overlap between the evolutionary tracks and the observational $\rm \Omega_{\star}$ seems to favour ages between $\rm 360~Myr$ and $\rm 3~Gyr$ for GJ 504, approximately, while no overlap is obtained for older ages. In Appendix~\ref{App: StarComp3MJu3.2Om}, we show the results obtained by doing a similar exercise, but this time we fix $\rm \Omega_{in} = 3.2~\Omega_{\odot}$, while letting the initial orbital distance vary between $\rm 0.02$ and $\rm 0.1~AU$. Analogously to what was found above, the shorter the initial orbital distance and the earlier the engulfment occurs. Moreover, also in this case the overlap between the tracks and the observational values tends to favour ages for GJ 504 up to $\rm \sim2.5~Gyr$. These results clearly show that despite the fact we manage to determine a region in $\rm M_{pl}$-$\rm a_{in}$-$\rm \Omega_{in}$ within which the theoretical tracks are compatible with the observational constraints for GJ 504, the problem still remains strongly degenerate. 

According to the two evolutionary scenarios analysed in this work only the ``Star with close-by companion'' is able to produce tracks compatible with both the surface rotation rate and X-ray luminosity of GJ 504, that results to be a strong candidate for a star that underwent strong star-planet interactions. It is interesting to notice that the planetary inward migration and consequent destruction at the Roche limit should have occurred late enough in the MS, otherwise the braking at the stellar surface due to magnetised winds would easily erase any trace of spin-up. 

It is worth stressing that despite our work supports the engulfment scenario as mean to reconcile the age estimations discrepancies between activity indicators and classical parameters, thus favouring a more evolved evolutionary stage for GJ 504, the problem of deriving accurate stellar parameters is still open. This will be crucial to finally establish the nature of GJ 504 and characterise the impact of SPI. In this context, asteroseismology is the only tool able to address this need. Future observations with TESS in the announced ultra-short cadence mode in the extended campaign might open to the opportunity of detecting solar-like oscillations.

\begin{acknowledgements}

CP thanks the Belgian Federal Science Policy Office (BELSPO) for the financial support in the framework of the PRODEX Program of the European Space Agency (ESA) under contract number 4000141194.

GB is funded by the Fonds National de la Recherche Scientifique (FNRS).

EM is supported by Deutsche Forschungsgemeinschaft under grant STE 1068/8-1. 

This work is based on data from eROSITA, the soft X-ray instrument aboard SRG, a joint Russian-German science mission supported by the Russian Space Agency (Roskosmos), in the interests of the Russian Academy of Sciences represented by its Space Research Institute (IKI), and the Deutsches Zentrum für Luft- und Raumfahrt (DLR). The SRG spacecraft was built by Lavochkin Association (NPOL) and its subcontractors, and is operated by NPOL with support from the Max Planck Institute for Extraterrestrial Physics (MPE).

The development and construction of the eROSITA X-ray instrument was led by MPE, with contributions from the Dr. Karl Remeis Observatory Bamberg \& ECAP (FAU Erlangen-Nuernberg), the University of Hamburg Observatory, the Leibniz Institute for Astrophysics Potsdam (AIP), and the Institute for Astronomy and Astrophysics of the University of Tübingen, with the support of DLR and the Max Planck Society. The Argelander Institute for Astronomy of the University of Bonn and the Ludwig Maximilians Universität Munich also participated in the science preparation for eROSITA.

The eROSITA data shown here were processed using the eSASS/NRTA software system developed by the German eROSITA consortium.

VVG is an F.R.S.-FNRS Research Associate.

SB acknowledges funding from the Dutch Research Council (NWO) with project number OCENW.M.22.215 of the research program `Open Competition Domain Science- M'.

\end{acknowledgements}

\bibliographystyle{aa}
\bibliography{biblioarticleGJ504}

\begin{thebibliography}{116}
\expandafter\ifx\csname natexlab\endcsname\relax\def\natexlab#1{#1}\fi

\bibitem[{{Adibekyan} {et~al.}(2018){Adibekyan}, {Sousa}, \&
  {Santos}}]{Adibekyan2018}
{Adibekyan}, V., {Sousa}, S.~G., \& {Santos}, N.~C. 2018, in Astrophysics and
  Space Science Proceedings, Vol.~49, Asteroseismology and Exoplanets:
  Listening to the Stars and Searching for New Worlds, ed. T.~L. {Campante},
  N.~C. {Santos}, \& M.~J.~P.~F.~G. {Monteiro}, 225

\bibitem[{{Arnaud}(1996)}]{Arnaud96.0}
{Arnaud}, K.~A. 1996, in Astronomical Society of the Pacific Conference Series,
  Vol. 101, Astronomical Data Analysis Software and Systems V, ed. G.~H.
  {Jacoby} \& J.~{Barnes}, 17

\bibitem[{{Asplund} {et~al.}(2009){Asplund}, {Grevesse}, {Sauval}, \&
  {Scott}}]{Asplund2009}
{Asplund}, M., {Grevesse}, N., {Sauval}, A.~J., \& {Scott}, P. 2009, \araa, 47,
  481

\bibitem[{{Barker}(2020)}]{Barker2020}
{Barker}, A.~J. 2020, \mnras, 498, 2270

\bibitem[{{Barker} \& {Ogilvie}(2010)}]{Barker2010}
{Barker}, A.~J. \& {Ogilvie}, G.~I. 2010, \mnras, 404, 1849

\bibitem[{{Barnes}(2007)}]{Barnes2007}
{Barnes}, S.~A. 2007, \apj, 669, 1167

\bibitem[{{Bashi} {et~al.}(2017){Bashi}, {Helled}, {Zucker}, \&
  {Mordasini}}]{Bashi2017}
{Bashi}, D., {Helled}, R., {Zucker}, S., \& {Mordasini}, C. 2017, \aap, 604,
  A83

\bibitem[{{Benbakoura} {et~al.}(2019){Benbakoura}, {R{\'e}ville}, {Brun}, {Le
  Poncin-Lafitte}, \& {Mathis}}]{Benbakoura2019}
{Benbakoura}, M., {R{\'e}ville}, V., {Brun}, A.~S., {Le Poncin-Lafitte}, C., \&
  {Mathis}, S. 2019, \aap, 621, A124

\bibitem[{{Benomar} {et~al.}(2018){Benomar}, {Bazot}, {Nielsen}, {Gizon},
  {Sekii}, {Takata}, {Hotta}, {Hanasoge}, {Sreenivasan}, \&
  {Christensen-Dalsgaard}}]{Benomar2018}
{Benomar}, O., {Bazot}, M., {Nielsen}, M.~B., {et~al.} 2018, Science, 361, 1231

\bibitem[{{Benomar} {et~al.}(2015){Benomar}, {Takata}, {Shibahashi},
  {Ceillier}, \& {Garc{\'\i}a}}]{Benomar2015}
{Benomar}, O., {Takata}, M., {Shibahashi}, H., {Ceillier}, T., \&
  {Garc{\'\i}a}, R.~A. 2015, \mnras, 452, 2654

\bibitem[{{B{\'e}trisey} {et~al.}(2023){B{\'e}trisey}, {Eggenberger},
  {Buldgen}, {Benomar}, \& {Bazot}}]{Betrisey2023}
{B{\'e}trisey}, J., {Eggenberger}, P., {Buldgen}, G., {Benomar}, O., \&
  {Bazot}, M. 2023, \aap, 673, L11

\bibitem[{{Bolmont} \& {Mathis}(2016)}]{BolmontMathis2016}
{Bolmont}, E. \& {Mathis}, S. 2016, Celestial Mechanics and Dynamical
  Astronomy, 126, 275

\bibitem[{{Bonnefoy} {et~al.}(2018){Bonnefoy}, {Perraut}, {Lagrange},
  {Delorme}, {Vigan}, {Line}, {Rodet}, {Ginski}, {Mourard}, {Marleau},
  {Samland}, {Tremblin}, {Ligi}, {Cantalloube}, {Molli{\`e}re}, {Charnay},
  {Kuzuhara}, {Janson}, {Morley}, {Homeier}, {D'Orazi}, {Klahr}, {Mordasini},
  {Lavie}, {Baudino}, {Beust}, {Peretti}, {Musso Bartucci}, {Mesa},
  {B{\'e}zard}, {Boccaletti}, {Galicher}, {Hagelberg}, {Desidera}, {Biller},
  {Maire}, {Allard}, {Borgniet}, {Lannier}, {Meunier}, {Desort}, {Alecian},
  {Chauvin}, {Langlois}, {Henning}, {Mugnier}, {Mouillet}, {Gratton}, {Brandt},
  {Mc Elwain}, {Beuzit}, {Tamura}, {Hori}, {Brandner}, {Buenzli}, {Cheetham},
  {Cudel}, {Feldt}, {Kasper}, {Keppler}, {Kopytova}, {Meyer}, {Perrot},
  {Rouan}, {Salter}, {Schmidt}, {Sissa}, {Zurlo}, {Wildi}, {Blanchard}, {De
  Caprio}, {Delboulb{\'e}}, {Maurel}, {Moulin}, {Pavlov}, {Rabou}, {Ramos},
  {Roelfsema}, {Rousset}, {Stadler}, {Rigal}, \& {Weber}}]{Bonnefoy2018}
{Bonnefoy}, M., {Perraut}, K., {Lagrange}, A.~M., {et~al.} 2018, \aap, 618, A63

\bibitem[{{Brunner} {et~al.}(2022){Brunner}, {Liu}, {Lamer}, {Georgakakis},
  {Merloni}, {Brusa}, {Bulbul}, {Dennerl}, {Friedrich}, {Liu}, {Maitra},
  {Nandra}, {Ramos-Ceja}, {Sanders}, {Stewart}, {Boller}, {Buchner}, {Clerc},
  {Comparat}, {Dwelly}, {Eckert}, {Finoguenov}, {Freyberg}, {Ghirardini},
  {Gueguen}, {Haberl}, {Kreykenbohm}, {Krumpe}, {Osterhage}, {Pacaud},
  {Predehl}, {Reiprich}, {Robrade}, {Salvato}, {Santangelo}, {Schrabback},
  {Schwope}, \& {Wilms}}]{Brunner22.0}
{Brunner}, H., {Liu}, T., {Lamer}, G., {et~al.} 2022, \aap, 661, A1

\bibitem[{{Caldiroli} {et~al.}(2022){Caldiroli}, {Haardt}, {Gallo}, {Spinelli},
  {Malsky}, \& {Rauscher}}]{Caldiroli2022}
{Caldiroli}, A., {Haardt}, F., {Gallo}, E., {et~al.} 2022, \aap, 663, A122

\bibitem[{{Chen} \& {Rogers}(2016)}]{Chen2016}
{Chen}, H. \& {Rogers}, L.~A. 2016, \apj, 831, 180

\bibitem[{{Cumming} {et~al.}(1999){Cumming}, {Marcy}, \&
  {Butler}}]{Cumming1999}
{Cumming}, A., {Marcy}, G.~W., \& {Butler}, R.~P. 1999, \apj, 526, 890

\bibitem[{{Cuntz} {et~al.}(2000){Cuntz}, {Saar}, \& {Musielak}}]{Cuntz2000}
{Cuntz}, M., {Saar}, S.~H., \& {Musielak}, Z.~E. 2000, \apjl, 533, L151

\bibitem[{{da Silva} {et~al.}(2012){da Silva}, {Porto de Mello}, {Milone}, {da
  Silva}, {Ribeiro}, \& {Rocha-Pinto}}]{daSilva2012}
{da Silva}, R., {Porto de Mello}, G.~F., {Milone}, A.~C., {et~al.} 2012, \aap,
  542, A84

\bibitem[{{Di Mauro} {et~al.}(2022){Di Mauro}, {Reda}, {Mathur}, {Garc{\'\i}a},
  {Buzasi}, {Corsaro}, {Benomar}, {Gonz{\'a}lez Cuesta}, {Stassun}, {Benatti},
  {D'Orazi}, {Giovannelli}, {Mesa}, \& {Nardetto}}]{DiMauro2022}
{Di Mauro}, M.~P., {Reda}, R., {Mathur}, S., {et~al.} 2022, \apj, 940, 93

\bibitem[{{Donahue} {et~al.}(1996){Donahue}, {Saar}, \&
  {Baliunas}}]{Donahue1996}
{Donahue}, R.~A., {Saar}, S.~H., \& {Baliunas}, S.~L. 1996, \apj, 466, 384

\bibitem[{{D'Orazi} {et~al.}(2017){D'Orazi}, {Desidera}, {Gratton}, {Lanza},
  {Messina}, {Andrievsky}, {Korotin}, {Benatti}, {Bonnefoy}, {Covino}, \&
  {Janson}}]{DOrazi2017}
{D'Orazi}, V., {Desidera}, S., {Gratton}, R.~G., {et~al.} 2017, \aap, 598, A19

\bibitem[{{Duguid} {et~al.}(2024){Duguid}, {de Vries}, {Lecoanet}, \&
  {Barker}}]{Duguid2024}
{Duguid}, C.~D., {de Vries}, N.~B., {Lecoanet}, D., \& {Barker}, A.~J. 2024,
  \apjl, 966, L14

\bibitem[{{Eggenberger} {et~al.}(2019){Eggenberger}, {Buldgen}, \&
  {Salmon}}]{Eggenberger2019a}
{Eggenberger}, P., {Buldgen}, G., \& {Salmon}, S.~J.~A.~J. 2019, \aap, 626, L1

\bibitem[{{Erkaev} {et~al.}(2007){Erkaev}, {Kulikov}, {Lammer}, {Selsis},
  {Langmayr}, {Jaritz}, \& {Biernat}}]{Erkaev2007}
{Erkaev}, N.~V., {Kulikov}, Y.~N., {Lammer}, H., {et~al.} 2007, \aap, 472, 329

\bibitem[{{Essick} \& {Weinberg}(2016)}]{Essick2016}
{Essick}, R. \& {Weinberg}, N.~N. 2016, \apj, 816, 18

\bibitem[{{Farnir} {et~al.}(2020){Farnir}, {Dupret}, {Buldgen}, {Salmon},
  {Noels}, {Pin{\c{c}}on}, {Pezzotti}, \& {Eggenberger}}]{Farnir2020}
{Farnir}, M., {Dupret}, M.~A., {Buldgen}, G., {et~al.} 2020, \aap, 644, A37

\bibitem[{{Fellay} {et~al.}(2023){Fellay}, {Pezzotti}, {Buldgen},
  {Eggenberger}, \& {Bolmont}}]{Fellay2023}
{Fellay}, L., {Pezzotti}, C., {Buldgen}, G., {Eggenberger}, P., \& {Bolmont},
  E. 2023, \aap, 669, A2

\bibitem[{{Fischer} {et~al.}(2014){Fischer}, {Marcy}, \&
  {Spronck}}]{Fischer2014}
{Fischer}, D.~A., {Marcy}, G.~W., \& {Spronck}, J. F.~P. 2014, \apjs, 210, 5

\bibitem[{{Foster} {et~al.}(2022){Foster}, {Poppenhaeger}, {Ilic}, \&
  {Schwope}}]{Foster2022}
{Foster}, G., {Poppenhaeger}, K., {Ilic}, N., \& {Schwope}, A. 2022, \aap, 661,
  A23

\bibitem[{{Fuhrmann} \& {Chini}(2015)}]{Fuhrmann2015}
{Fuhrmann}, K. \& {Chini}, R. 2015, \apj, 806, 163

\bibitem[{{Fuller}(2017)}]{Fuller2017}
{Fuller}, J. 2017, \mnras, 472, 1538

\bibitem[{{Fuller} {et~al.}(2015){Fuller}, {Cantiello}, {Stello}, {Garcia}, \&
  {Bildsten}}]{Fuller2015}
{Fuller}, J., {Cantiello}, M., {Stello}, D., {Garcia}, R.~A., \& {Bildsten}, L.
  2015, Science, 350, 423

\bibitem[{{Garc{\'\i}a} {et~al.}(2007){Garc{\'\i}a}, {Turck-Chi{\`e}ze},
  {Jim{\'e}nez-Reyes}, {Ballot}, {Pall{\'e}}, {Eff-Darwich}, {Mathur}, \&
  {Provost}}]{Garcia2007}
{Garc{\'\i}a}, R.~A., {Turck-Chi{\`e}ze}, S., {Jim{\'e}nez-Reyes}, S.~J.,
  {et~al.} 2007, Science, 316, 1591

\bibitem[{{Garraffo} {et~al.}(2016){Garraffo}, {Drake}, \&
  {Cohen}}]{Garraffo2016}
{Garraffo}, C., {Drake}, J.~J., \& {Cohen}, O. 2016, \aap, 595, A110

\bibitem[{{Goldreich} \& {Nicholson}(1977)}]{Goldreich1977}
{Goldreich}, P. \& {Nicholson}, P.~D. 1977, \icarus, 30, 301

\bibitem[{{Goodman} \& {Dickson}(1998)}]{Goodman1998}
{Goodman}, J. \& {Dickson}, E.~S. 1998, \apj, 507, 938

\bibitem[{{Holmberg} {et~al.}(2009){Holmberg}, {Nordstr{\"o}m}, \&
  {Andersen}}]{Holmberg2009}
{Holmberg}, J., {Nordstr{\"o}m}, B., \& {Andersen}, J. 2009, \aap, 501, 941

\bibitem[{{Iglesias} \& {Rogers}(1996)}]{IglesiasRogers1996}
{Iglesias}, C.~A. \& {Rogers}, F.~J. 1996, \apj, 464, 943

\bibitem[{{Ili{\'c}} {et~al.}(2023){Ili{\'c}}, {Poppenhaeger}, {Dsouza},
  {Wolk}, {Ag{\"u}eros}, \& {Stelzer}}]{Ilic2023}
{Ili{\'c}}, N., {Poppenhaeger}, K., {Dsouza}, D., {et~al.} 2023, \mnras, 524,
  5954

\bibitem[{{Ili{\'c}} {et~al.}(2024){Ili{\'c}}, {Poppenhaeger}, {Queiroz}, \&
  {Chiappini}}]{Ilic2024}
{Ili{\'c}}, N., {Poppenhaeger}, K., {Queiroz}, A.~B., \& {Chiappini}, C. 2024,
  Astronomische Nachrichten, 345, e20230132

\bibitem[{{Irwin}(2012)}]{Irwin2012}
{Irwin}, A.~W. 2012, {FreeEOS: Equation of State for stellar interiors
  calculations}, Astrophysics Source Code Library, record ascl:1211.002

\bibitem[{{Ivanov} {et~al.}(2013){Ivanov}, {Papaloizou}, \&
  {Chernov}}]{Ivanov2013}
{Ivanov}, P.~B., {Papaloizou}, J.~C.~B., \& {Chernov}, S.~V. 2013, \mnras, 432,
  2339

\bibitem[{{Jin} {et~al.}(2014){Jin}, {Mordasini}, {Parmentier}, {van Boekel},
  {Henning}, \& {Ji}}]{Jin2014}
{Jin}, S., {Mordasini}, C., {Parmentier}, V., {et~al.} 2014, \apj, 795, 65

\bibitem[{{Johnstone} {et~al.}(2021){Johnstone}, {Bartel}, \&
  {G{\"u}del}}]{Johnstone2021}
{Johnstone}, C.~P., {Bartel}, M., \& {G{\"u}del}, M. 2021, \aap, 649, A96

\bibitem[{{Kozai}(1962)}]{Kozai1962}
{Kozai}, Y. 1962, \aj, 67, 591

\bibitem[{{Kraft}(1967)}]{Kraft1967}
{Kraft}, R.~P. 1967, \apj, 150, 551

\bibitem[{{Kubyshkina} {et~al.}(2018){Kubyshkina}, {Fossati}, {Erkaev},
  {Cubillos}, {Johnstone}, {Kislyakova}, {Lammer}, {Lendl}, \&
  {Odert}}]{Kubyshkina2018}
{Kubyshkina}, D., {Fossati}, L., {Erkaev}, N.~V., {et~al.} 2018, \apjl, 866,
  L18

\bibitem[{{Kubyshkina} \& {Fossati}(2021)}]{Kubyshkina2021}
{Kubyshkina}, D.~I. \& {Fossati}, L. 2021, Research Notes of the American
  Astronomical Society, 5, 74

\bibitem[{{Kuzuhara} {et~al.}(2013){Kuzuhara}, {Tamura}, {Kudo}, {Janson},
  {Kandori}, {Brandt}, {Thalmann}, {Spiegel}, {Biller}, {Carson}, {Hori},
  {Suzuki}, {Burrows}, {Henning}, {Turner}, {McElwain}, {Moro-Mart{\'\i}n},
  {Suenaga}, {Takahashi}, {Kwon}, {Lucas}, {Abe}, {Brandner}, {Egner}, {Feldt},
  {Fujiwara}, {Goto}, {Grady}, {Guyon}, {Hashimoto}, {Hayano}, {Hayashi},
  {Hayashi}, {Hodapp}, {Ishii}, {Iye}, {Knapp}, {Matsuo}, {Mayama}, {Miyama},
  {Morino}, {Nishikawa}, {Nishimura}, {Kotani}, {Kusakabe}, {Pyo}, {Serabyn},
  {Suto}, {Takami}, {Takato}, {Terada}, {Tomono}, {Watanabe}, {Wisniewski},
  {Yamada}, {Takami}, \& {Usuda}}]{Kuzuhara2013}
{Kuzuhara}, M., {Tamura}, M., {Kudo}, T., {et~al.} 2013, \apj, 774, 11

\bibitem[{{Lammer} {et~al.}(2003){Lammer}, {Selsis}, {Ribas}, {Guinan},
  {Bauer}, \& {Weiss}}]{Lammer2003}
{Lammer}, H., {Selsis}, F., {Ribas}, I., {et~al.} 2003, \apjl, 598, L121

\bibitem[{{Lazovik}(2021)}]{Lazovik2021}
{Lazovik}, Y.~A. 2021, \mnras, 508, 3408

\bibitem[{{Lebreton} \& {Reese}(2020)}]{Lebreton2020}
{Lebreton}, Y. \& {Reese}, D.~R. 2020, \aap, 642, A88

\bibitem[{{Lecoanet} {et~al.}(2022){Lecoanet}, {Bowman}, \& {Van
  Reeth}}]{Lecoanet2022}
{Lecoanet}, D., {Bowman}, D.~M., \& {Van Reeth}, T. 2022, \mnras, 512, L16

\bibitem[{{Lecoanet} {et~al.}(2017){Lecoanet}, {Vasil}, {Fuller}, {Cantiello},
  \& {Burns}}]{Lecoanet2017}
{Lecoanet}, D., {Vasil}, G.~M., {Fuller}, J., {Cantiello}, M., \& {Burns},
  K.~J. 2017, \mnras, 466, 2181

\bibitem[{{Lidov}(1962)}]{Lidov1962}
{Lidov}, M.~L. 1962, \planss, 9, 719

\bibitem[{{Lopez} \& {Fortney}(2014)}]{Lopez2014}
{Lopez}, E.~D. \& {Fortney}, J.~J. 2014, \apj, 792, 1

\bibitem[{{Ma} \& {Fuller}(2021)}]{MaFuller2021}
{Ma}, L. \& {Fuller}, J. 2021, \apj, 918, 16

\bibitem[{{Magaudda} {et~al.}(2022){Magaudda}, {Stelzer}, {Raetz}, {Klutsch},
  {Salvato}, \& {Wolf}}]{Magaudda22.0}
{Magaudda}, E., {Stelzer}, B., {Raetz}, S., {et~al.} 2022, \aap, 661, A29

\bibitem[{{Maggio} {et~al.}(2007){Maggio}, {Flaccomio}, {Favata}, {Micela},
  {Sciortino}, {Feigelson}, \& {Getman}}]{Maggio07.0}
{Maggio}, A., {Flaccomio}, E., {Favata}, F., {et~al.} 2007, \apj, 660, 1462

\bibitem[{{Mathis}(2015)}]{Mathis2015}
{Mathis}, S. 2015, \aap, 580, L3

\bibitem[{{Matt} {et~al.}(2015){Matt}, {Brun}, {Baraffe}, {Bouvier}, \&
  {Chabrier}}]{Matt2015}
{Matt}, S.~P., {Brun}, A.~S., {Baraffe}, I., {Bouvier}, J., \& {Chabrier}, G.
  2015, \apjl, 799, L23

\bibitem[{{Matt} {et~al.}(2019){Matt}, {Brun}, {Baraffe}, {Bouvier}, \&
  {Chabrier}}]{Matt2019}
{Matt}, S.~P., {Brun}, A.~S., {Baraffe}, I., {Bouvier}, J., \& {Chabrier}, G.
  2019, \apjl, 870, L27

\bibitem[{{Mazeh} {et~al.}(1997){Mazeh}, {Krymolowski}, \&
  {Rosenfeld}}]{Mazeh1997}
{Mazeh}, T., {Krymolowski}, Y., \& {Rosenfeld}, G. 1997, \apjl, 477, L103

\bibitem[{{Merloni} {et~al.}(2024){Merloni}, {Larmer}, {Liu}, {Ramos-Ceja},
  {Brunner}, {Bubul}, {Dennerli}, {Doroshenko}, {Freyberg}, {Friedrich},
  {Gatuzz}, {Georgakakis}, {Haberl}, {Igo}, {Kreykenbohm}, {Liu}, {Maitra},
  {Malyali}, {Mayer}, {Nandra}, {Predehl}, {Robrade}, {Salvato}, {Sanders},
  {Stewart}, {Tubín-Arenas}, {Weber}, {Wilms}, {Arcodia}, {Artis},
  {Aschersleben}, {Avakyan}, {Aydar}, {Bahar}, {Balzer}, {Becker}, {Berger},
  {Boller}, {Bornemann}, {Brüggen}, {Brusa}, {Buchner}, {Burwitz},
  {Camilloni}, {Clerc}, {Comparat}, {Coutinho}, {Czesla}, {Dannhauer},
  {Dauner}, {Dauser}, {Dietl}, {Dolag}, {Dwelly}, {Egg}, {Ehl}, {Freund},
  {Friedrich}, {Gaida}, {Garrel}, {Ghirardini}, {Gokus}, {Grünwald},
  {Grandis}, {Grotova}, {Gruen}, {Gueguen}, {Hämmerich}, {Hamaus}, {Hasinger},
  {Haubner}, {Homan}, {IderChitham}, {Joseph}, {Joyce}, {König},
  {Kaltenbrunner}, {Khokhriakova}, {Kink}, {Kirsch}, {Kluge}, {Knies},
  {Krippendorf}, {Krumpe}, {Kurpas}, {Li}, {Liu}, {Locatelli}, {Lorenz},
  {Müller}, {Magaudda}, {Mannes}, {McCall}, {Meidinger}, {Michailidis},
  {Migkas}, {Muñoz-Giraldo}, {Musiimenta}, {Nguyen-Dang}, {Ni}, {Olechowska},
  {Ota}, {Pacaud}, {Pasini}, {Perinati}, {Pires}, {Pommranz}, {Ponti},
  {Poppenhaeger}, {Pühlhofer}, {Rau}, {Reh}, {Reiprich}, {Roster}, {Saeedi},
  {Santangelo}, {Sasaki}, {Schmitt}, {Schneider}, {Schrabback}, {Schuster},
  {Schwope}, {Seppi}, {Serim}, {Shreeram}, {Sokolova-Lapa}, {Starck},
  {Stelzer}, {Stierhof}, {Suleimanov}, {Tenzer}, {Traulsen}, {Trümper},
  {Tsuge}, {Urrutia}, {Veronica}, {Waddell}, {Willer}, {Wolf}, {Yeung},
  {Zainab}, {Zangrandi}, {Zhang}, {Zhang}, \& {Zheng}}]{Merloni24.0}
{Merloni}, A., {Larmer}, G., {Liu}, T., {et~al.} 2024, \aap, 682, A34

\bibitem[{{Messina} {et~al.}(2003){Messina}, {Pizzolato}, {Guinan}, \&
  {Rodon{\`o}}}]{Messina2003}
{Messina}, S., {Pizzolato}, N., {Guinan}, E.~F., \& {Rodon{\`o}}, M. 2003,
  \aap, 410, 671

\bibitem[{{Metcalfe} {et~al.}(2024){Metcalfe}, {Strassmeier}, {Ilyin},
  {Buzasi}, {Kochukhov}, {Ayres}, {Basu}, {Chontos}, {Finley}, {See},
  {Stassun}, {van Saders}, {Sepulveda}, \& {Ricker}}]{Metcalfe2024}
{Metcalfe}, T.~S., {Strassmeier}, K.~G., {Ilyin}, I.~V., {et~al.} 2024, \apjl,
  960, L6

\bibitem[{{Metzger} {et~al.}(2012){Metzger}, {Giannios}, \&
  {Spiegel}}]{Metzger2012}
{Metzger}, B.~D., {Giannios}, D., \& {Spiegel}, D.~S. 2012, \mnras, 425, 2778

\bibitem[{{Miglio} \& {Montalb{\'a}n}(2005)}]{Miglio2005}
{Miglio}, A. \& {Montalb{\'a}n}, J. 2005, \aap, 441, 615

\bibitem[{{Nielsen} {et~al.}(2015){Nielsen}, {Schunker}, {Gizon}, \&
  {Ball}}]{Nielsen2015}
{Nielsen}, M.~B., {Schunker}, H., {Gizon}, L., \& {Ball}, W.~H. 2015, \aap,
  582, A10

\bibitem[{{{\'O} Fionnag{\'a}in} \& {Vidotto}(2018)}]{OFionnagain2018}
{{\'O} Fionnag{\'a}in}, D. \& {Vidotto}, A.~A. 2018, \mnras, 476, 2465

\bibitem[{{Oetjens} {et~al.}(2020){Oetjens}, {Carone}, {Bergemann}, \&
  {Serenelli}}]{Oetjens2020}
{Oetjens}, A., {Carone}, L., {Bergemann}, M., \& {Serenelli}, A. 2020, \aap,
  643, A34

\bibitem[{{Ogilvie}(2013)}]{Ogilvie2013}
{Ogilvie}, G.~I. 2013, \mnras, 429, 613

\bibitem[{{Ogilvie} \& {Lin}(2004)}]{Ogilvie2004}
{Ogilvie}, G.~I. \& {Lin}, D.~N.~C. 2004, \apj, 610, 477

\bibitem[{{Ogilvie} \& {Lin}(2007)}]{Ogilvie2007}
{Ogilvie}, G.~I. \& {Lin}, D.~N.~C. 2007, \apj, 661, 1180

\bibitem[{{Otegi} {et~al.}(2020){Otegi}, {Bouchy}, \& {Helled}}]{Otegi2020}
{Otegi}, J.~F., {Bouchy}, F., \& {Helled}, R. 2020, \aap, 634, A43

\bibitem[{{Owen} \& {Wu}(2013)}]{Owen2013}
{Owen}, J.~E. \& {Wu}, Y. 2013, \apj, 775, 105

\bibitem[{{Parker}(1955)}]{Parker1955}
{Parker}, E.~N. 1955, \apj, 122, 293

\bibitem[{{Pezzotti} {et~al.}(2021){Pezzotti}, {Eggenberger}, {Buldgen},
  {Meynet}, {Bourrier}, \& {Mordasini}}]{Pezzotti2021}
{Pezzotti}, C., {Eggenberger}, P., {Buldgen}, G., {et~al.} 2021, \aap, 650,
  A108

\bibitem[{{Pillitteri} {et~al.}(2022){Pillitteri}, {Micela}, {Maggio},
  {Sciortino}, \& {Lopez-Santiago}}]{Pillitteri2022}
{Pillitteri}, I., {Micela}, G., {Maggio}, A., {Sciortino}, S., \&
  {Lopez-Santiago}, J. 2022, \aap, 660, A75

\bibitem[{{Pont} \& {Eyer}(2004)}]{Pont2004}
{Pont}, F. \& {Eyer}, L. 2004, \mnras, 351, 487

\bibitem[{{Poppenhaeger} \& {Wolk}(2014)}]{Poppenhaeger2014}
{Poppenhaeger}, K. \& {Wolk}, S.~J. 2014, \aap, 565, L1

\bibitem[{{Predehl} {et~al.}(2021){Predehl}, {Andritschke}, {Arefiev},
  {Babyshkin}, {Batanov}, {Becker}, {B{\"o}hringer}, {Bogomolov}, {Boller},
  {Borm}, {Bornemann}, {Br{\"a}uninger}, {Br{\"u}ggen}, {Brunner}, {Brusa},
  {Bulbul}, {Buntov}, {Burwitz}, {Burkert}, {Clerc}, {Churazov}, {Coutinho},
  {Dauser}, {Dennerl}, {Doroshenko}, {Eder}, {Emberger}, {Eraerds},
  {Finoguenov}, {Freyberg}, {Friedrich}, {Friedrich}, {F{\"u}rmetz},
  {Georgakakis}, {Gilfanov}, {Granato}, {Grossberger}, {Gueguen}, {Gureev},
  {Haberl}, {H{\"a}lker}, {Hartner}, {Hasinger}, {Huber}, {Ji}, {Kienlin},
  {Kink}, {Korotkov}, {Kreykenbohm}, {Lamer}, {Lomakin}, {Lapshov}, {Liu},
  {Maitra}, {Meidinger}, {Menz}, {Merloni}, {Mernik}, {Mican}, {Mohr},
  {M{\"u}ller}, {Nandra}, {Nazarov}, {Pacaud}, {Pavlinsky}, {Perinati},
  {Pfeffermann}, {Pietschner}, {Ramos-Ceja}, {Rau}, {Reiffers}, {Reiprich},
  {Robrade}, {Salvato}, {Sanders}, {Santangelo}, {Sasaki}, {Scheuerle},
  {Schmid}, {Schmitt}, {Schwope}, {Shirshakov}, {Steinmetz}, {Stewart},
  {Str{\"u}der}, {Sunyaev}, {Tenzer}, {Tiedemann}, {Tr{\"u}mper}, {Voron},
  {Weber}, {Wilms}, \& {Yaroshenko}}]{Predehl2021}
{Predehl}, P., {Andritschke}, R., {Arefiev}, V., {et~al.} 2021, \aap, 647, A1

\bibitem[{{Privitera} {et~al.}(2016){Privitera}, {Meynet}, {Eggenberger},
  {Vidotto}, {Villaver}, \& {Bianda}}]{Privitera2016B}
{Privitera}, G., {Meynet}, G., {Eggenberger}, P., {et~al.} 2016, \aap, 593,
  A128

\bibitem[{{Rao} {et~al.}(2018){Rao}, {Meynet}, {Eggenberger}, {Haemmerl{\'e}},
  {Privitera}, {Georgy}, {Ekstr{\"o}m}, \& {Mordasini}}]{Rao2018}
{Rao}, S., {Meynet}, G., {Eggenberger}, P., {et~al.} 2018, \aap, 618, A18

\bibitem[{{Rao} {et~al.}(2021){Rao}, {Pezzotti}, {Meynet}, {Eggenberger},
  {Buldgen}, {Mordasini}, {Bourrier}, {Ekstr{\"o}m}, \& {Georgy}}]{Rao2021}
{Rao}, S., {Pezzotti}, C., {Meynet}, G., {et~al.} 2021, \aap, 651, A50

\bibitem[{{Rasio} \& {Ford}(1996)}]{RasioFord1996}
{Rasio}, F.~A. \& {Ford}, E.~B. 1996, Science, 274, 954

\bibitem[{{Rasio} {et~al.}(1996){Rasio}, {Tout}, {Lubow}, \&
  {Livio}}]{Rasio1996}
{Rasio}, F.~A., {Tout}, C.~A., {Lubow}, S.~H., \& {Livio}, M. 1996, \apj, 470,
  1187

\bibitem[{{R{\'e}ville} {et~al.}(2015){R{\'e}ville}, {Brun}, {Matt},
  {Strugarek}, \& {Pinto}}]{Reville2015}
{R{\'e}ville}, V., {Brun}, A.~S., {Matt}, S.~P., {Strugarek}, A., \& {Pinto},
  R.~F. 2015, \apj, 798, 116

\bibitem[{{Ricker} {et~al.}(2014){Ricker}, {Winn}, {Vanderspek}, {Latham},
  {Bakos}, {Bean}, {Berta-Thompson}, {Brown}, {Buchhave}, {Butler}, {Butler},
  {Chaplin}, {Charbonneau}, {Christensen-Dalsgaard}, {Clampin}, {Deming},
  {Doty}, {De Lee}, {Dressing}, {Dunham}, {Endl}, {Fressin}, {Ge}, {Henning},
  {Holman}, {Howard}, {Ida}, {Jenkins}, {Jernigan}, {Johnson}, {Kaltenegger},
  {Kawai}, {Kjeldsen}, {Laughlin}, {Levine}, {Lin}, {Lissauer}, {MacQueen},
  {Marcy}, {McCullough}, {Morton}, {Narita}, {Paegert}, {Palle}, {Pepe},
  {Pepper}, {Quirrenbach}, {Rinehart}, {Sasselov}, {Sato}, {Seager},
  {Sozzetti}, {Stassun}, {Sullivan}, {Szentgyorgyi}, {Torres}, {Udry}, \&
  {Villasenor}}]{Ricker2014}
{Ricker}, G.~R., {Winn}, J.~N., {Vanderspek}, R., {et~al.} 2014, in Society of
  Photo-Optical Instrumentation Engineers (SPIE) Conference Series, Vol. 9143,
  Space Telescopes and Instrumentation 2014: Optical, Infrared, and Millimeter
  Wave, ed. J.~{Oschmann}, Jacobus~M., M.~{Clampin}, G.~G. {Fazio}, \& H.~A.
  {MacEwen}, 914320

\bibitem[{{Rui} \& {Fuller}(2023)}]{Rui2023}
{Rui}, N.~Z. \& {Fuller}, J. 2023, \mnras, 523, 582

\bibitem[{{Saio} {et~al.}(2021){Saio}, {Takata}, {Lee}, {Li}, \& {Van
  Reeth}}]{Saio2021}
{Saio}, H., {Takata}, M., {Lee}, U., {Li}, G., \& {Van Reeth}, T. 2021, \mnras,
  502, 5856

\bibitem[{{Salz} {et~al.}(2016){Salz}, {Schneider}, {Czesla}, \&
  {Schmitt}}]{Salz2016}
{Salz}, M., {Schneider}, P.~C., {Czesla}, S., \& {Schmitt}, J.~H.~M.~M. 2016,
  \aap, 585, L2

\bibitem[{{Scuflaire} {et~al.}(2008){Scuflaire}, {Th{\'e}ado}, {Montalb{\'a}n},
  {Miglio}, {Bourge}, {Godart}, {Thoul}, \& {Noels}}]{Scuflaire2008}
{Scuflaire}, R., {Th{\'e}ado}, S., {Montalb{\'a}n}, J., {et~al.} 2008, \apss,
  316, 83

\bibitem[{{Shkolnik} {et~al.}(2003){Shkolnik}, {Walker}, \&
  {Bohlender}}]{Shkolnik2003}
{Shkolnik}, E., {Walker}, G.~A.~H., \& {Bohlender}, D.~A. 2003, \apj, 597, 1092

\bibitem[{{Skumanich}(1972)}]{Skumanich1972}
{Skumanich}, A. 1972, \apj, 171, 565

\bibitem[{{Soderblom}(2010)}]{Soderblom2010}
{Soderblom}, D.~R. 2010, \araa, 48, 581

\bibitem[{{Soderblom} {et~al.}(2014){Soderblom}, {Hillenbrand}, {Jeffries},
  {Mamajek}, \& {Naylor}}]{Soderblom2014}
{Soderblom}, D.~R., {Hillenbrand}, L.~A., {Jeffries}, R.~D., {Mamajek}, E.~E.,
  \& {Naylor}, T. 2014, in Protostars and Planets VI, ed. H.~{Beuther}, R.~S.
  {Klessen}, C.~P. {Dullemond}, \& T.~{Henning}, 219--241

\bibitem[{{Strugarek}(2024)}]{Strugarek2024}
{Strugarek}, A. 2024, Comptes Rendus Physique, 24, 138

\bibitem[{{Sunyaev} {et~al.}(2021){Sunyaev}, {Arefiev}, {Babyshkin},
  {Bogomolov}, {Borisov}, {Buntov}, {Brunner}, {Burenin}, {Churazov},
  {Coutinho}, {Eder}, {Eismont}, {Freyberg}, {Gilfanov}, {Gureyev}, {Hasinger},
  {Khabibullin}, {Kolmykov}, {Komovkin}, {Krivonos}, {Lapshov}, {Levin},
  {Lomakin}, {Lutovinov}, {Medvedev}, {Merloni}, {Mernik}, {Mikhailov},
  {Molodtsov}, {Mzhelsky}, {M{\"u}ller}, {Nandra}, {Nazarov}, {Pavlinsky},
  {Poghodin}, {Predehl}, {Robrade}, {Sazonov}, {Scheuerle}, {Shirshakov},
  {Tkachenko}, \& {Voron}}]{Sunyaev2021}
{Sunyaev}, R., {Arefiev}, V., {Babyshkin}, V., {et~al.} 2021, \aap, 656, A132

\bibitem[{{Takeda} {et~al.}(2007){Takeda}, {Ford}, {Sills}, {Rasio}, {Fischer},
  \& {Valenti}}]{Takeda2007}
{Takeda}, G., {Ford}, E.~B., {Sills}, A., {et~al.} 2007, \apjs, 168, 297

\bibitem[{{Truemper}(1982)}]{Trumper1982}
{Truemper}, J. 1982, Advances in Space Research, 2, 241

\bibitem[{{Valenti} \& {Fischer}(2005)}]{Valenti2005}
{Valenti}, J.~A. \& {Fischer}, D.~A. 2005, \apjs, 159, 141

\bibitem[{{van Saders} {et~al.}(2016){van Saders}, {Ceillier}, {Metcalfe},
  {Silva Aguirre}, {Pinsonneault}, {Garc{\'\i}a}, {Mathur}, \&
  {Davies}}]{VanSaders2016}
{van Saders}, J.~L., {Ceillier}, T., {Metcalfe}, T.~S., {et~al.} 2016, \nat,
  529, 181

\bibitem[{{Vernazza} {et~al.}(1981){Vernazza}, {Avrett}, \&
  {Loeser}}]{Vernazza1981}
{Vernazza}, J.~E., {Avrett}, E.~H., \& {Loeser}, R. 1981, \apjs, 45, 635

\bibitem[{{Vidotto}(2020)}]{Vidotto2020}
{Vidotto}, A.~A. 2020, in Solar and Stellar Magnetic Fields: Origins and
  Manifestations, ed. A.~{Kosovichev}, S.~{Strassmeier}, \& M.~{Jardine}, Vol.
  354, 259--267

\bibitem[{{Villaver} \& {Livio}(2009)}]{Villaver2009}
{Villaver}, E. \& {Livio}, M. 2009, \apjl, 705, L81

\bibitem[{{Voges} {et~al.}(1999){Voges}, {Aschenbach}, {Boller},
  {Br{\"a}uninger}, {Briel}, {Burkert}, {Dennerl}, {Englhauser}, {Gruber},
  {Haberl}, {Hartner}, {Hasinger}, {K{\"u}rster}, {Pfeffermann}, {Pietsch},
  {Predehl}, {Rosso}, {Schmitt}, {Tr{\"u}mper}, \& {Zimmermann}}]{Voges1999}
{Voges}, W., {Aschenbach}, B., {Boller}, T., {et~al.} 1999, \aap, 349, 389

\bibitem[{{Watson} {et~al.}(1981){Watson}, {Donahue}, \& {Walker}}]{Watson1981}
{Watson}, A.~J., {Donahue}, T.~M., \& {Walker}, J.~C.~G. 1981, \icarus, 48, 150

\bibitem[{{Weinberg} {et~al.}(2012){Weinberg}, {Arras}, {Quataert}, \&
  {Burkart}}]{Weinberg2012}
{Weinberg}, N.~N., {Arras}, P., {Quataert}, E., \& {Burkart}, J. 2012, \apj,
  751, 136

\bibitem[{{Wilson}(1966)}]{Wilson1966}
{Wilson}, O.~C. 1966, \apj, 144, 695

\bibitem[{{Wright} {et~al.}(2011){Wright}, {Drake}, {Mamajek}, \&
  {Henry}}]{Wright2011}
{Wright}, N.~J., {Drake}, J.~J., {Mamajek}, E.~E., \& {Henry}, G.~W. 2011,
  \apj, 743, 48

\bibitem[{{Wright} {et~al.}(2018){Wright}, {Newton}, {Williams}, {Drake}, \&
  {Yadav}}]{Wright2018}
{Wright}, N.~J., {Newton}, E.~R., {Williams}, P. K.~G., {Drake}, J.~J., \&
  {Yadav}, R.~K. 2018, \mnras, 479, 2351

\bibitem[{{Zahn}(1966)}]{Zahn1966}
{Zahn}, J.~P. 1966, Annales d'Astrophysique, 29, 489

\bibitem[{{Zahn}(1977)}]{Zahn1977}
{Zahn}, J.~P. 1977, \aap, 500, 121

\bibitem[{{Zhang} \& {Penev}(2014)}]{ZhangPenev2014}
{Zhang}, M. \& {Penev}, K. 2014, \apj, 787, 131

\end{thebibliography}

\newpage

\begin{appendix}

\section{Star with $\rm 1~M_{J}$ companion}
\label{App: StarComp1MJu}

\begin{figure}
\includegraphics[width=\linewidth]{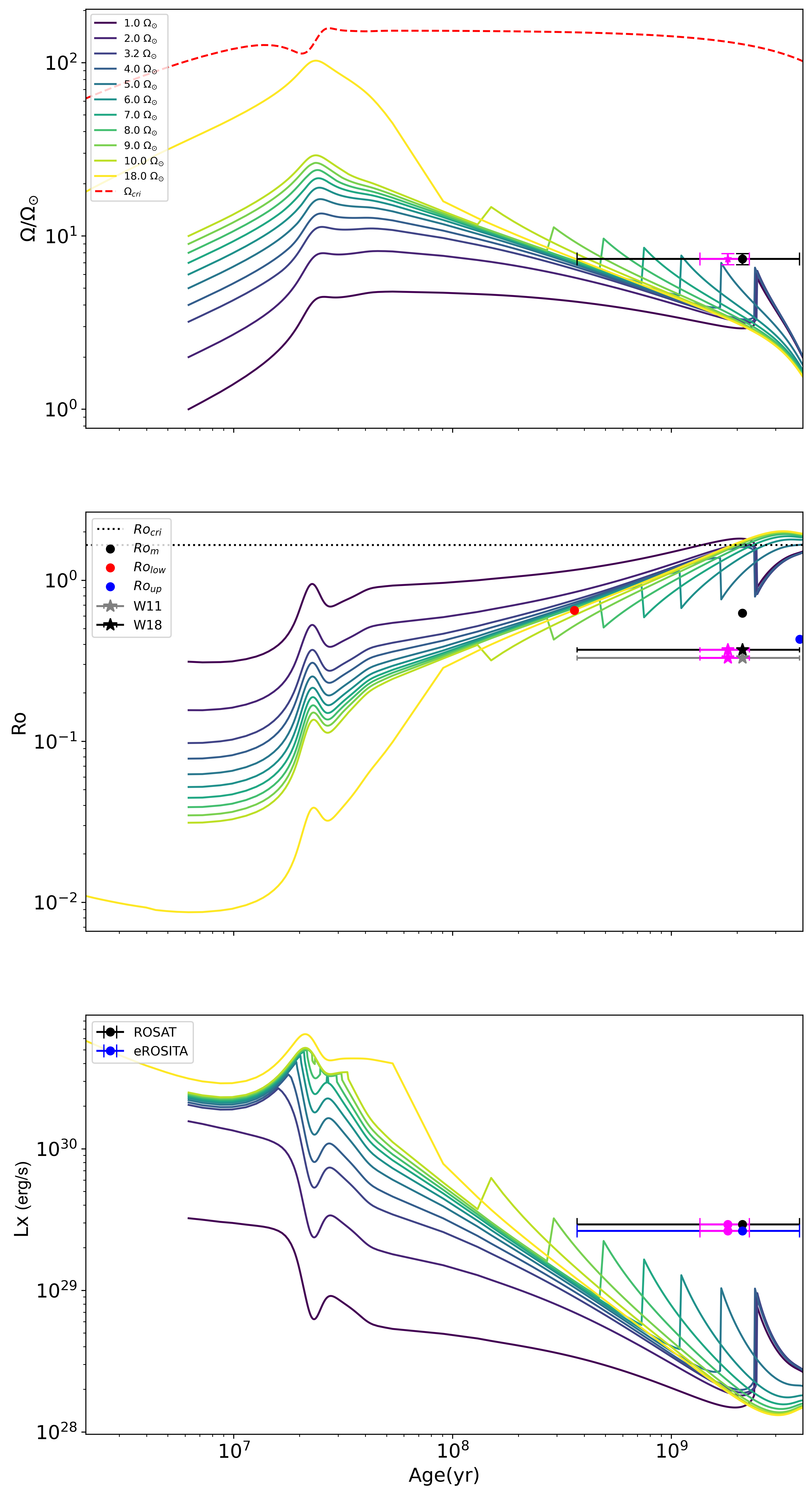}
\caption{\small{Evolution of the stellar surface rotation rate (\emph{Top panel}), stellar Rossby number (\emph{Middle panel}) and X-ray luminosity (\emph{Bottom panel}), under the impact of planetary inward migration. For the stellar initial surface rotation rate we considered the following values $\rm \Omega_{in} = 1, 2, 3.2, 4, 5, 6, 7, 8, 9, 10, 18~\Omega_{\odot}$, for the planetary mass $\rm M_{pl} = 1~M_{J}$, and initial planetary orbital distance $\rm a_{in} = 0.02~AU$. The meaning of the other quantities is the same as in Fig.~\ref{Fig: Single Star}.}}
\label{Fig: Star companion 1MJu}
\end{figure}

The evolution of the host star surface rotation rate and X-ray luminosity is significantly affected by a planetary companion at $\rm a_{in} = 0.02~AU$, and with initial mass $\rm 1~M_{J}$. As we show in Fig.~\ref{Fig: Star companion 1MJu}, the star spins-up when the planet migrates inwards and reaches its Roche limit, and simultaneously its X-ray emission increases. Nevertheless, a $\rm 1~M_{J}$ planet is not massive enough to transfer the required angular momentum from its orbit to the star to increase its X-ray luminosity at the point to be compatible with the observational data. The minimum mass required, according to our computations, is $\rm 2~M_{J}$.

\section{Star with $\rm 3~M_{J}$ companion and $\rm \Omega_{in} = 3.2~\Omega_{\odot}$}
\label{App: StarComp3MJu3.2Om}

\begin{figure}
\includegraphics[width=\linewidth]{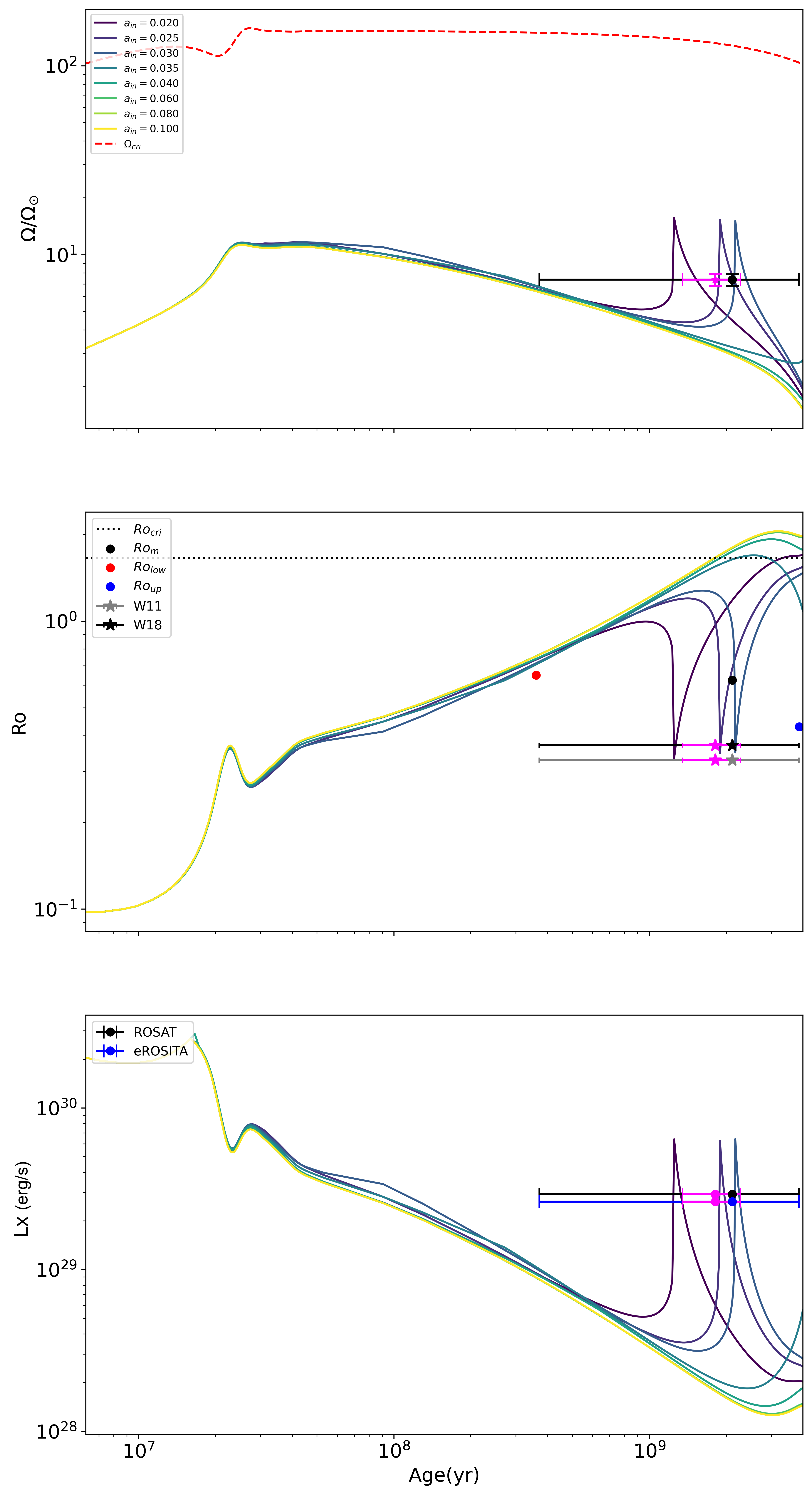}
 \caption{\small{Evolution of the stellar surface rotation rate (\emph{Top panel}), stellar Rossby number (\emph{Middle panel}) and X-ray luminosity (\emph{Bottom panel}), under the impact of planetary inward migration. In this case, we fix the initial mass of the planet ($\rm3~M_{J}$), the initial surface rotation rate of the host star ($\rm \Omega_{in} = 3.2~\Omega_{\odot}$), while we vary the initial orbital distance ($\rm a_{in} = 0.02,0.025,0.03,0.035,0.04, 0.06, 0.08, 0.1~AU$). The meaning of the other quantities is the same as in Fig.~\ref{Fig: Single Star}.}}
 \label{Fig: Star companion 3MJu 3.2Om}
\end{figure}

In Fig.~\ref{Fig: Star companion} we show how the host star surface rotation rate and X-ray luminosity are impacted by the inward migration of a putative planet with $\rm M_{in} = 3~M_{J}$ and $\rm a_{in} = 0.025~AU$, while for the host star a range of initial surface rotation rates is explored ($\rm \Omega_{in} = 1, 2, 3.2, 4, 5, 6, 7, 8, 9, 10, 18~\Omega_{\odot}$). Similarly, in Fig.~\ref{Fig: Star companion 3MJu 3.2Om} we show the results obtained by fixing $\rm \Omega_{in}$ to $\rm 3.2~\Omega_{\odot}$ and varying the initial orbital distance in the range $\rm a_{in} = 0.02,0.025,0.03,0.035,0.04, 0.06, 0.08, 0.1~AU$. In this context, the shorter the orbital distance and the earlier the engulfment of the planet occurs.

\end{appendix}

\end{document}